\documentclass[lettersize,journal]{IEEEtran}

\usepackage{amsmath,amsfonts}
\usepackage{algorithmic}
\usepackage{algorithm}
\usepackage{array}
\usepackage{mathtools}
\usepackage{textcomp}
\usepackage{stfloats}
\usepackage{url}
\usepackage{verbatim}
\usepackage{graphicx}
\usepackage{cite}
\usepackage{glossaries}
\usepackage{xcolor}
\usepackage{makecell}
\usepackage{booktabs}
\usepackage{multirow}
\usepackage{tikz}
\usepackage{fancyhdr}
\usepackage[position=b]{subcaption}
\usepackage[hidelinks]{hyperref}
\usepackage[nameinlink]{cleveref}
\usepackage{pifont}
\usepackage{orcidlink}
\usepackage{threeparttable}
\definecolor{color_ls}{RGB}{151, 180, 104}
\definecolor{color_mc}{RGB}{242, 149,  69}
\definecolor{color_rd}{RGB}{248,  46, 190}
\definecolor{color_cp}{RGB}{218, 118, 113}
\definecolor{cmark_green}{RGB}{0,150,0}
\definecolor{xmark_red}{RGB}{180,0,0}
\newif\ifanonymous
\anonymousfalse

\newcommand{\anon}[2]{%
  \ifanonymous
    #1%
  \else
    #2%
  \fi
}

\newcommand{\cmark}{\textcolor{cmark_green}{\ding{51}}}%
\newcommand{\xmark}{\textcolor{xmark_red}{\ding{55}}}%

\Crefname{figure}{Fig.}{Figs.}

\makeatletter
\newcommand{\insertnewlines}[1]{%
  \noindent\mbox{}%
  \@tempcnta=#1\relax
  \loop\ifnum\@tempcnta>0
    \\
    \advance\@tempcnta\m@ne
  \repeat
}
\makeatother

\newacronym{gsm}{GSM}{Graduate Student Member}
\newacronym{rtl}{RTL}{Register Transfer Level}

\newacronym[plural=WANs, firstplural={Wide Area Networks (WANs)}]{wan}{WAN}{Wide Area Network}
\newacronym[plural=WSNs, firstplural={Wireless Sensor Networks (WSNs)}]{wsn}{WSN}{Wireless Sensor Network}
\newacronym{simd}{SIMD}{Single Instruction Multiple Data}
\newacronym{os}{OS}{Operating System}
\newacronym{ble}{BLE}{Bluetooth Low-Energy}
\newacronym{wifi}{Wi-FI}{Wireless Fidelity}
\newacronym[plural=DVS, firstplural={Dynamic Vision Sensors (DVS)}]{dvs}{DVS}{Dynamic Vision Sensor}
\newacronym{ptz}{PTZ}{Pan-Tilt Unit}

\newacronym[plural=FLLs,firstplural=Frequency Locked Loops (FLLs)]{fll}{FLL}{Frequency Locked Loop}
\newacronym{dram}{DRAM}{Dynamic Random Access Memory}
\newacronym{fpu}{FPU}{Floating Point Unit}
\newacronym{fpss}{FPSS}{Floating Point Subsystem}
\newacronym{frep}{FREP}{Floating Point Repetition}
\newacronym{dma}{DMA}{Direct Memory Access}
\newacronym{dca}{DCA}{Direct Compute Access}
\newacronym{ssr}{SSR}{Stream Semantic Register}
\newacronym{issr}{ISSR}{Indirection Stream Semantic Register}
\newacronym[plural=LUTs, firstplural={Lookup Tables (LUTs)}]{lut}{LUT}{Lookup Table}
\newacronym[plural=FPGAs, firstplural={Field Programmable Gate Arrays (FPGAs)}]{fpga}{FPGA}{Field Programmable Gate Array}
\newacronym{dsp}{DSP}{Digital Signal Processing}
\newacronym{mcu}{MCU}{Microcontroller Unit}
\newacronym{spi}{SPI}{Serial Peripheral Interface}
\newacronym{cpi}{CPI}{Camera Parallel Interface}
\newacronym{rf}{RF}{Register File}
\newacronym{fifo}{FIFO}{First-In First-Out Queue}
\newacronym{uart}{UART}{Universal Asynchronous Receiver-Transmitter}
\newacronym{raw}{RAW}{Read After Write}
\newacronym[plural=ISAs, firstplural={Instruction Set Architectures (ISAs)}]{isa}{ISA}{Instruction Set Architecture}
\newacronym{xbar}{XBAR}{crossbar}
\newacronym[firstplural=Scratch-Pad Memories (SPMs)]{spm}{SPM}{Scratch-Pad Memory}
\newacronym{ppa}{PPA}{Power Performance Area}
\newacronym{ipi}{IPI}{Inter-Processor Interrupt}
\newacronym[firstplural=Software-Generated Interrupts (SGIs)]{sgi}{SGI}{Software-Generated Interrupt}
\newacronym{pe}{PE}{Processing Element}
\newacronym{tcdm}{TCDM}{Tightly-Coupled Data Memory}
\newacronym{lsu}{LSU}{Load-Store Unit}
\newacronym{icache}{I\$}{Instruction Cache}
\newacronym{dcache}{D\$}{Data Cache}
\newacronym{wfi}{WFI}{Wait For Interrupt}
\newacronym{gpc}{GPC}{GPU Processing Cluster}
\newacronym{cpu}{CPU}{Central Processing Unit}
\newacronym{gpu}{GPU}{Graphics Processing Unit}
\newacronym{llc}{LLC}{Last-Level Cache}
\newacronym{sm}{SM}{Streaming Multiprocessor}
\newacronym[firstplural=Networks on Chip (NoCs)]{noc}{NoC}{Network on Chip}
\newacronym{ni}{NI}{Network Interface}
\newacronym[firstplural=Gate Equivalents (GEs)]{ge}{GE}{Gate Equivalent}
\newacronym{hbm}{HBM}{High Bandwidth Memory}

\newacronym{dfg}{DFG}{Data Flow Graph}
\newacronym{lcg}{LCG}{Linear Congruential Generator}
\newacronym{prn}{PRN}{Pseudo-Random Number}
\newacronym{gemm}{GEMM}{General Matrix Multiplication}

\newacronym{ste}{STE}{Straight-Through-Estimator}

\newacronym[plural=PTUs, firstplural={Pan-Tilt Units}]{ptu}{PTU}{Pan-Tilt Unit}
\newacronym{mdf}{MDF}{Medium-density fibreboard}
\newacronym{cvat}{CVAT}{Computer Vision Annotation Tool}
\newacronym{coco}{COCO}{Common Objects in Context}
\newacronym{soa}{SoA}{State-of-the-Art}
\newacronym{sf}{SF}{Sensor Fusion}
\newacronym{mtp}{MTP}{Multi-Token Prediction}

\newacronym{dl}{DL}{Deep Learning}
\newacronym{bn}{BN}{Batch Normalization}
\newacronym{FGSM}{FBK}{Fast Gradient Sign Method}
\newacronym{lr}{LR}{Learning Rate}
\newacronym{sgd}{SGD}{Stochastic Gradient Descent}
\newacronym{gd}{GD}{Gradient Descent}
\newacronym{llm}{LLM}{Large Language Model}

\newacronym{sta}{STA}{Static Timing Analysis}

\newacronym[plural=GPIOs, firstplural={General Purpose Inupt Outputs (GPIOs)}]{gpio}{GPIO}{General Purpose Input Output}
\newacronym[plural=LDOs, firstplural={Low Dropout Regulators (LDOs)}]{ldo}{LDO}{Low Dropout Regulator}

\newacronym{inq}{INQ}{Incremental Network Quantization}

\newacronym{CV}{CV}{Computer Vision}
\newacronym{EoT}{EoT}{Expectation over Transformation}
\newacronym{RPN}{RPN}{Region Proposal Network}
\newacronym{TV}{TV}{Total Variation}
\newacronym{NPS}{NPS}{Non-Printability Score}
\newacronym{STN}{STN}{Spatial Transformer Network}
\newacronym{MTCNN}{MTCNN}{Multi-Task Convolutional Neural Network}
\newacronym{YOLO}{YOLO}{You Only Look Once}
\newacronym{SSD}{SSD}{Single Shot Detector}
\newacronym{SOTA}{SOTA}{State of the Art}
\newacronym{NMS}{NMS}{Non-Maximum Suppression}
\newacronym{ic}{IC}{Integrated Circuit}
\newacronym{tcxo}{TCXO}{Temperature Controlled Crystal Oscillator}
\newacronym{jtag}{JTAG}{Joint Test Action Group industry standard}
\newacronym{swd}{SWD}{Serial Wire Debug}
\newacronym{sdio}{SDIO}{Serial Data Input Output}

\newacronym[plural=PCBs, firstplural={Printed Circuit Boards (PCB)}]{pcb}{PCB}{Printed Circuit Board}
\newacronym[plural=ASICs, firstplural={Application Specific Integrated Circuits}]{asic}{ASIC}{Application Specific Integrated Circuit}

\newacronym[plural=BNNs, firstplural={Binary Neural Networks (BNNs)}]{bnn}{BNN}{Binary Neural Network}
\newacronym[plural=NNs, firstplural={Neural Networks}]{nn}{NN}{Neural Network (NNs)}
\newacronym[plural=SCMs, firstplural={Standard Cell Memories (SCMs)}]{scm}{SCM}{Standard Cell Memory}
\newacronym{ann}{ANN}{Artificial Neural Networks}
\newacronym{ml}{ML}{Machine Learning}
\newacronym{ai}{AI}{Artificial Intelligence}
\newacronym{iot}{IoT}{Internet of Things}
\newacronym{fft}{FFT}{Fast Fourier Transform}
\newacronym[plural=OCUs, firstplural={Output Channel Compute Units (OCUs)}]{ocu}{OCU}{Output Channel Compute Unit}
\newacronym{alu}{ALU}{Arithmetic Logic Unit}
\newacronym{mac}{MAC}{Multiply-Accumulate}
\newacronym[firstplural={systems-on-chip (SoCs)}]{soc}{SoC}{system-on-chip}
\newacronym[firstplural={multi-processor systems-on-chip (MPSoCs)}]{mpsoc}{MPSoC}{multi-processor system-on-chip}

\newacronym{PGD}{PGD}{Projected Gradient Descend}
\newacronym{CW}{CW}{Carlini-Wagner}
\newacronym{OD}{OD}{Object Detection}

\newacronym{rrf}{RRF}{RADAR Repetition Frequency}
\newacronym{nlp}{NLP}{Natural Language Processing}
\newacronym{qam}{QAM}{Quadrature Amplitude Modulation}
\newacronym{rri}{RRI}{RADAR Repetition Interval}
\newacronym{radar}{RADAR}{Radio Detection and Ranging}
\newacronym{loocv}{LOOCV}{Leave-one-out cross validation}

\newacronym{bsp}{BSP}{Board Support Package}
\newacronym{ttn}{TTN}{The Things Network}
\newacronym{wip}{WIP}{Work in Progress}
\newacronym{json}{JSON}{JavaScript Object Notation}
\newacronym{qat}{QAT}{Quantization-Aware Training}

\newacronym{cls}{CLS}{Classification Error}
\newacronym{loc}{LOC}{Localization Error}
\newacronym{bkgd}{BKGD}{Background Error}
\newacronym{roc}{ROC}{Receiver Operating Characteristic}
\newacronym{frr}{FRR}{False Rejection Rate}
\newacronym{eer}{EER}{Equal Error Rate}
\newacronym{snr}{SNR}{Signal-to-Noise Ratio}
\newacronym{flop}{FLOP}{Floating-Point Operation}
\newacronym{fp}{FP}{Floating-Point}
\newacronym{fps}{FPS}{Frames Per Second}
\newacronym{oi}{OI}{Operational Intensity}
\newacronym{ipc}{IPC}{Instructions per Cycle}
\newacronym{tdp}{TDP}{Thermal Design Power}

\newacronym{gsc}{GSC}{Google Speech Commands}
\newacronym{mswc}{MSWC}{Multilingual Spoken Words Corpus}
\newacronym{demand}{DEMAND}{Diverse Environments Multichannel Acoustic Noise Database}

\newacronym[plural=SNNs, firstplural={Spiking Neural Networks (SNNs)}]{snn}{SNN}{Spiking Neural Network}
\newacronym[plural=DNNs, firstplural={Deep Neural Networks (DNNs)}]{dnn}{DNN}{Deep Neural Network}
\newacronym[plural=TCNs,firstplural=Temporal Convolutional Networks]{tcn}{TCN}{Temporal Convolutional Network}
\newacronym[plural=CNNs,firstplural=Convolutional Neural Networks (CNNs)]{cnn}{CNN}{Convolutional Neural Network}
\newacronym[plural=TNNs,firstplural=Ternarized Neural Networks]{tnn}{TNN}{Ternarized Neural Network}
\newacronym{ds-cnn}{DS-CNN}{Depthwise Separable Convolutional Neural Network}
\newacronym{rnn}{RNN}{Recurrent Neural Network}
\newacronym{gcn}{GCN}{Graph Convolutional Network}
\newacronym{mha}{MHA}{Multi-Head Attention}
\newacronym{mqa}{MQA}{Multi-Query Attention}
\newacronym{gqa}{GQA}{Grouped-Query Attention}
\newacronym{mla}{MLA}{Multi-Head Latent Attention}
\newacronym{ffn}{FFN}{Feed-Forward Network}
\newacronym{mlp}{MLP}{Multilayer Perceptron}
\newacronym{moe}{MoE}{Mixture of Experts}
\newacronym{kv}{KV}{Key-Value}
\newacronym[shortplural=RoPE]{rope}{RoPE}{Rotary Position Embedding}
\newacronym{silu}{SiLU}{Sigmoid-weighted Linear Unit}
\newacronym{crnn}{CRNN}{Convolutional Recurrent Neural Network}
\newacronym{clca}{CLCA}{Convolutional Linear Cross-Attention}

\newacronym{bf}{BF}{Beamforming}
\newacronym{anc}{ANC}{Active Noise Cancellation}
\newacronym{agc}{AGC}{Automatic Gain Control}
\newacronym{se}{SE}{Speech Enhancement}
\newacronym{mct}{MCT}{Multi-Condition Training}
\newacronym{mcta}{MCTA}{Multi-Condition Training \& Adaptation}
\newacronym{pcen}{PCEN}{Per-Channel Energy Normalization}
\newacronym{mfcc}{MFCC}{Mel-Frequency Cepstral Coefficient}
\newacronym{asr}{ASR}{Automated Speech Recognition}
\newacronym{kws}{KWS}{Keyword Spotting}
\newacronym{odl}{ODL}{On-Device Learning}

\newacronym{nl-kws}{NL-KWS}{Noiseless Keyword Spotting}
\newacronym{na-kws}{NA-KWS}{Noise-Aware Keyword Spotting}
\newacronym{odda}{ODDA}{On-Device Domain Adaptation}
\newacronym{hpm}{HPM}{High-Performance Mode}
\newacronym{lpm}{LPM}{Low-Power Mode}

\newacronym{tpot}{TPOT}{Time per Output Token}
\newacronym{pp}{PP}{Pipeline Parallelism}
\newacronym{ep}{EP}{Expert Parallelism}
\newacronym{tp}{TP}{Tensor Parallelism}
\newacronym{dp}{DP}{Data Parallelism}
\newacronym{sp}{SP}{Sequence Parallelism}
\newacronym{c2c}{C2C}{Chip-to-Chip}
\newacronym{d2d}{D2D}{Die-to-Die}

\newacronym{rvv}{RVV}{RISC-V Vector}
\newacronym{ce}{CE}{Compute Element}

\begin{document}

\title{FlatAttention: Dataflow and Fabric Collectives Co-Optimization for Large Attention-Based Model Inference on Tile-Based Accelerators}

\ifanonymous
\author{Anonymous\orcidlink{0000-0000-0000-000},~\IEEEmembership{Member,~IEEE,}
}
\else
\author{Chi~Zhang\orcidlink{0009-0003-3243-0558},~\IEEEmembership{Graduate Student Member (GSM),~IEEE,}
        Luca~Colagrande\orcidlink{0000-0002-7986-1975},~\IEEEmembership{GSM,~IEEE,}
        Renzo~Andri\orcidlink{0000-0002-8776-5158},~\IEEEmembership{Member,~IEEE,}
        and Luca Benini\orcidlink{0000-0001-8068-3806},~\IEEEmembership{Fellow,~IEEE}
\thanks{\textbf{This work has been submitted to the IEEE for possible publication.
Copyright may be transferred without notice, after which this version may no longer be accessible.}}
\thanks{Chi Zhang, Luca Colagrande, and Luca Benini are with the Integrated Systems Laboratory (IIS), ETH Zurich, 8092 Zurich, Switzerland (e-mail: chizhang@iis.ee.ethz.ch; colluca@iis.ee.ethz.ch; lbenini@iis.ee.ethz.ch).}
\thanks{Luca Benini is also with the Department of Electrical, Electronic and Information Engineering (DEI), University of Bologna, 40126 Bologna, Italy}
\thanks{Renzo Andri is with the Computing Systems Laboratory, Huawei Technologies Switzerland AG, Zurich, Switzerland (e-mail: renzo.andri@huawei.com).}
        }%
\fi

\newcommand{\ARCH}{Tile-Based Many-PE }
\newcommand{\arch}{tile-based many-PE }
\newcommand{\SYS}{Wafer-scale Multi-die System }
\newcommand{\sys}{wafer-scale multi-die system }
\newcommand{\syspl}{wafer-scale multi-die systems }
\newcommand{\ResAMulticastSuSwseq}{30.7}
\newcommand{\ResAMulticastSuSwtree}{5.1}
\newcommand{\ResARedsumSuSwseq}{67.3}
\newcommand{\ResARedsumSuSwtree}{10.9}
\newcommand{\ResAFlashBW}{80}
\newcommand{\ResAFlatBW}{1.7}
\newcommand{\ResAFlatCollBW}{8.1}
\newcommand{\ResAOptFlatBW}{22.1}
\newcommand{\ResAOptFlatBWSU}{3.8}
\newcommand{\ResAOptFlatHBMTraficSave}{16}
\newcommand{\ResAOptFlatPerfSU}{4.1}
\newcommand{\ResAMaxUti}{92.3}

\newcommand{\ResBUtia}{92.7}
\newcommand{\ResBUtib}{92.3}
\newcommand{\ResBUtiMlow}{23}
\newcommand{\ResBRedMuleBestUti}{98}
\newcommand{\ResBFlatAvgUti}{86}
\newcommand{\ResBFlatAvgBWUti}{78}
\newcommand{\ResBFlatGPUSU}{1.9}

\newcommand{\ResCGEMMAvgSU}{1.9}

\newcommand{\ResDDSFlatSU}{2.1}
\newcommand{\ResDDSSoAPerfSUNvGPU}{2.9}
\newcommand{\ResDDSSoATpotSUNvGPU}{1.4}
\newcommand{\ResDDSSoAPerfSUDSProf}{1.9}
\newcommand{\ResDDSSoAPeakRedDSProf}{1.5}

\maketitle

\begin{abstract}
Attention accounts for an increasingly dominant fraction of total computation during inference for mixture-of-experts (MoE) models, making efficient acceleration critical.
Emerging domain-specific accelerators for large model inference are shifting toward chip-scale and wafer-scale tile-based architectures. 
Tiles contain large matrix and vector engines and are connected through on-chip interconnects, which support tile-to-tile traffic to reduce the tile-to-main-memory traffic bottleneck. Hence, dataflow management is crucial to achieve high utilization.
We propose FlatAttention, a dataflow for modern attention variants on tile-based accelerators.
FlatAttention minimizes expensive high-bandwidth memory (HBM) accesses by exploiting collective primitives integrated into the on-chip network fabric, achieving up to \ResAMaxUti\% utilization, \ResAOptFlatPerfSU$\times$ speedup over FlashAttention-3, and \ResAOptFlatHBMTraficSave$\times$ lower HBM traffic. On a 32×32 tile configuration with peak performance comparable to NVIDIA GH200, FlatAttention generalizes across multiple attention variants, achieving an average of \ResBFlatAvgUti\% utilization for compute-bound attentions and \ResBFlatAvgBWUti\% HBM bandwidth utilization for memory-bound ones, resulting in an average \ResBFlatGPUSU$\times$ speedup over attention implementations on GH200.
Finally, we evaluate end-to-end DeepSeek-v3 FP8 decoding with FlatAttention on a wafer-scale multi-die system, achieving a \ResDDSSoAPerfSUDSProf$\times$ improvement in system throughput and a \ResDDSSoATpotSUNvGPU$\times$ reduction in per-user token output latency, despite operating with \ResDDSSoAPeakRedDSProf$\times$ lower peak system performance compared to the state-of-the-art solution.

\end{abstract}

\begin{IEEEkeywords}
Large Language Models, Multi-Head Attention, Tile-Based Architecture, Network on Chip, Collective Primitives.
\end{IEEEkeywords}

\section{Introduction}
\label{sec:intro}

\begin{figure*}[t]
    \centering
    \begin{subfigure}{\columnwidth}
        \centering
        \includegraphics[width=\textwidth]{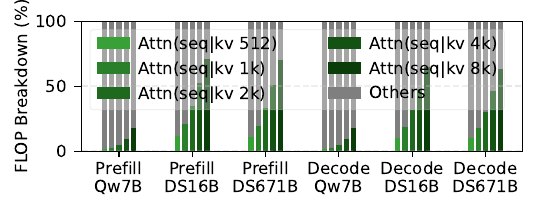}
        \put(-230,13){\small (a)}
        \captionsetup{labelformat=empty}
        \caption{}
        \vspace{-1.6em}
        \label{fig:flop-breakdown}
    \end{subfigure}
    \hfill
    \begin{subfigure}{\columnwidth}
        \centering
        \includegraphics[width=\textwidth]{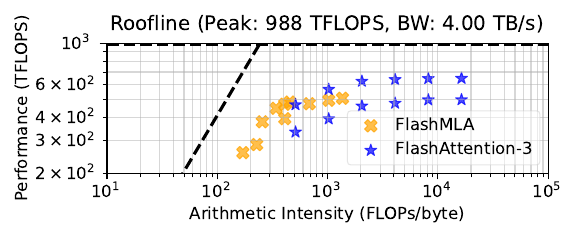}
        \put(-230,13){\small (b)}
        \captionsetup{labelformat=empty}
        \caption{}
        \vspace{-1.6em}
        \label{fig:gpu_not_good}
    \end{subfigure}
    \vspace{-0.6em}
    \caption{(a) FLOP breakdown for LLM models during prefill (\textit{seq} length) and decode (\textit{kv} length) stages. (b) Roofline plot of FlashAttention-3
    prefill and FlashMLA decode performance on Nvidia GH200 GPU. Evaluated with FP16 precision, varying head dimension and sequence length for prefill, while varying speculative length and KV cache length for decoding \anon{\cite{benchmarkresultsano}}{\cite{benchmarkresults}}.}
    \label{fig_flop_breakdown}
    \vspace{-0.5em}
\end{figure*}

\IEEEPARstart{T}{ransformer}-based \gls{ai} models, such as GPT-4, LLaMA, and DeepSeek-v3, have gained tremendous importance and are now a key, dominant workload for an increasing number of AI-serving systems. As these models continue to grow in size and complexity, they require substantial computational resources to achieve real-time performance. Consequently, efficient and scalable \gls{llm} inference has become a key challenge, driving innovations in both model architecture evolution and hardware acceleration solutions.

Recent models have moved from the classical decoder design, featuring stacked \gls{mha} and \gls{mlp} layers as in GPT and BERT \cite{vaswani2017}, to more efficient attention variants such as \gls{mqa} \cite{shazeer2019fast} and \gls{gqa} \cite{ainslie2023gqa} in the LLaMA family, and further to advanced mechanisms like \gls{mla} combined with \gls{moe} in DeepSeek-v3 \cite{deepseekv3}.
As \gls{moe}-based models significantly reduce the compute cost of the \gls{ffn} component, the attention mechanism accounts for an increasingly dominant fraction of total compute.
Reflecting this trend, \Cref{fig:flop-breakdown} compares the floating-point operation breakdown of the attention mechanism and other computational kernels between the \gls{mha}+\gls{mlp} model Qwen-chat-7B (\emph{Qw7B}) and \gls{mla}+\gls{moe} models DeepSeek-v3-16B (\emph{DS16B}) and DeepSeek-v3-671B (\emph{DS671B}), during both the prefill and decode stages.
At long-context inference, the attention mechanism of \emph{Qw7B} accounts for 19\% of all floating-point operations, whereas in \emph{DS671B} this proportion increases to 71\% during decoding, with a similar trend observed in the prefill stage.

Such ``attention bottleneck'' has led to extensive research aimed at optimizing attention dataflows on the dominant \gls{ai} hardware platform, namely Nvidia \glspl{gpu}.
One of the most widely adopted solutions is FlashAttention \cite{dao2022} for \gls{mha} dataflow, which accelerates attention by efficiently fusing its microkernels.
Over two generations of improvements, FlashAttention-2 \cite{dao2023} introduces further algorithmic optimizations, while FlashAttention-3 \cite{shah2024} leverages the asynchronous execution capabilities of the latest Nvidia \glspl{gpu}.
At the same time, recent application trends, driven by the increasing importance of ``reasoning'' models, have shifted focus toward inference efficiency. This has led to the development of attention variants such as \gls{mqa}, \gls{gqa} and \gls{mla}, aiming to reduce KV cache size and accelerate attention during inference on \glspl{gpu}.
An optimized dataflow for \gls{mla} on GPUs is FlashMLA \cite{deepseek2025flashmla}, developed by the DeepSeek team, which leverages a mechanism similar to FlashAttention.

Despite these advances, attention kernels still achieve only a fraction of the peak performance offered by the \gls{soa} Nvidia GH200 \gls{gpu}. \Cref{fig:gpu_not_good} shows the performance of FlashAttention-3 during the prefill stage and FlashMLA during decoding on the GH200 roofline model. Both implementations exhibit a large performance gap relative to the roofline, ranging from 26\% to 64\%.
Moreover, Nvidia’s GH200 \gls{gpu} comes with significant cost and power requirements: it integrates an 814 mm\textsuperscript{2} die fabricated in TSMC's 5nm process, coupled with six \gls{hbm} stacks that account for more than half of the total cost, adding up to a total \gls{tdp} of 700 W.

\begin{figure*}[t!]
  \centering
  \includegraphics[width=1.0\linewidth, trim={0.7cm 0.2cm 0.7cm 0.2cm}, clip]{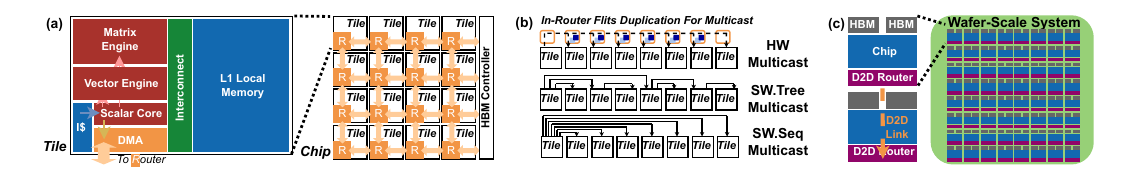}
  \begingroup
    \phantomsubcaption
    \label{fig_arch}
    \phantomsubcaption
    \label{fig_coll_demo}
    \phantomsubcaption
    \label{fig_multi_die_arch}
  \endgroup
  \vspace{-6mm}
  \caption{(a) Tile-based many-PE architecture template (b) Row-wise multicast implementation with fabric-supported hardware collectives (\textit{HW}) compared against two software-based collective implementations (\textit{SW.Tree} and \textit{SW.Seq}) (c) A \sys consisting of multiple tile-based many-\gls{pe} accelerators, with a 2D-mesh \gls{d2d} interconnect topology.}
  \label{fig_background}
\end{figure*}

Given the suboptimal utilization and the high cost and power requirements of modern \glspl{gpu}, industry and academia are aggressively developing accelerators for \gls{llm} inference, aiming to deliver competitive performance while boosting the efficiency of the system by minimizing energy-hungry \gls{hbm} accesses, in hope to significantly reduce per-token cost, while maintaining flexibility and competitive token serving latency and throughput.
\Cref{fig_arch} illustrates an emerging scalable design pattern for inference accelerators targeting large transformer-based \gls{ai} models \cite{liao2021, prabhakar2024b, lie2024, vasiljevic2024}. These systems adopt large, full-reticle, multi-die or even wafer-scale architectures \cite{lie2024}, structured as meshes of compute tiles that integrate thousands of \glspl{pe}, as extremely dense, large matrix processing units alongside vector and scalar engines to accelerate all major \gls{llm} kernels.
Each tile is provided with a local, software-managed L1 scratchpad memory to buffer data and hide main memory latency.
Multiple \glspl{hbm} are typically employed as main memory, positioned at die boundaries and connected to dedicated memory controllers.

Mapping \gls{llm} workloads onto these architectures presents a significant challenge.
While full, fine-grained control over inter-tile and tile-to-\gls{hbm} transfers enables highly optimized data movement, this flexibility results in large design space: the dataflow must be carefully designed to maximize matrix engine utilization while minimizing costly off-chip \gls{hbm} accesses.
Furthermore, co-designing a tile-based accelerator template that can efficiently map \gls{llm} workloads remains an open architectural problem which is tightly coupled with dataflow selection.

Among various parallel computing paradigms, collective communication—such as multicast and reduction—plays a crucial role in enabling efficient data exchange among processing tiles.
Intensively researched and widely used in distributed systems, these operations are now being adopted in tile-based many-PE architectures, especially with the introduction of hardware-supported collective primitives on the \gls{noc} \cite{metamtia2023,prabhakar2024b,vasiljevic2024}.
Traditional software-based collective primitives rely on successive point-to-point inter-tile transfers, resulting in high communication latency.
In contrast, \glspl{noc} equipped with hardware-supported collective communication primitives establish direct, optimized communication paths, significantly reducing communication overhead.
Recent work has proposed dedicated hardware implementations of fabric-level collective primitives, demonstrating high performance with modest area and power overheads \cite{colagrande2026,vasiljevic2024,deng2025torrent,prabhakar2024b,lie2022cerebras}.
These advances motivate the co-design of dataflow with \gls{noc}-level collective primitives, enabling improved on-chip data reuse, reduced HBM traffic, and higher \gls{pe} utilization.

Furthermore, modern LLMs continue to scale from tens to hundreds—and even thousands—of billions of parameters \cite{deepseekv3,fedus2022}.
As a result, LLM inference serving increasingly relies on interconnecting multiple accelerator chips and HBM stacks to meet the growing demands for memory capacity and compute throughput, as a single accelerator is fundamentally constrained by reticle size limits and the slowing of Moore’s Law.
Recent advances in wafer-scale technology, which enable high-density integration of multiple dies, have emerged as a promising system-integration paradigm for high-performance LLM serving \cite{he2025waferllm,xu2025wsc}.
By placing chips in close physical proximity and interconnecting them with high-speed \gls{d2d} links, wafer-scale systems significantly improve communication efficiency while reducing system integration overhead.
This trend raises another key question: \emph{``how does accelerating the attention bottleneck on a single accelerator die translate to end-to-end modern LLM inference performance across multiple dies, and in wafer-scale systems?''}

In this work, we address these challenges in a bottom-up manner. We first propose \textit{FlatAttention}, a dataflow that efficiently maps attention kernels and its variants on tile-based accelerators by leveraging collective communication primitives, such as multicast and reduction operations, in the \gls{noc}, to dramatically reduce off-chip memory accesses.
We demonstrate that co-designing the architecture to natively accelerate collective communication primitives enables mapping attention and its variants more efficiently than on \gls{soa} GPU solutions.
Furthermore, we take the full DeepSeek-v3-671B inference as an end-to-end study, mapping the complete decoding workload onto a \sys composed of tile-based many-\gls{pe} accelerators. With the use of \textit{FlatAttention}, we achieve substantial end-to-end decoding speedups over \gls{soa} solutions.
the contribuions of this work are:
\begin{itemize}
    \item Propose FlatAttention, an efficient workload allocation and scheduling strategy for modern attention variants like \gls{mha}, \gls{gqa} and \gls{mla} on tile-based accelerators.
    For evaluation, we developed a modeling and simulation framework to estimate the performance of a wide range of tile-based accelerator configurations, calibrated against cycle-accurate \gls{rtl} simulations.
    FlatAttention leverages collective primitives in the on-chip network fabric to achieve up to \ResAMaxUti\% utilization for the \gls{mha} layer on tile-based many-\gls{pe} accelerators, delivering a 4.1$\times$ speedup over the FlashAttention-3 dataflow on the same tile-based accelerator while reducing \gls{hbm} traffic by 16$\times$.
    \item Co-explore the accelerator architecture and FlatAttention parameters, identifying key trends and tradeoffs to guide the selection of optimal algorithm–architecture configurations.
    We propose a general tiling strategy for FlatAttention dataflow and identify an optimal tile-based accelerator configuration that matches the peak FP16 performance and \gls{hbm} bandwidth of NVIDIA’s state-of-the-art GH200 \gls{gpu}.
    By generalizing FlatAttention to advanced attention variants like \gls{gqa} and \gls{mla}, FlatAttention on this tile-based accelerator configuration achieves an average utilization of 86\% (up to 95.6\%) for attention kernels in the compute-bound regime, and an average \gls{hbm} bandwidth utilization of 78\% (up to 92.1\%) for memory-bound attention.
    Overall, FlatAttention on the selected architecture delivers an average $1.9\times$ speedup over optimized FlashAttention-3 and FlashMLA implementations on GH200, across both prefill and decode phases and multiple attention variants.
    \item Evaluate FlatAttention in an end-to-end deployment of DeepSeek-v3-671B decoder, a \gls{soa} open-source \gls{llm}, on a \sys composed of tile-based many-\gls{pe} accelerators.
    By examining various parallelism paradigms for multi-chip systems, we identify a highly optimized mapping of the DeepSeek-v3-671B decoder in terms of both overall throughput and per-user output token latency.
    Our approach achieves a \ResDDSSoAPerfSUDSProf$\times$ improvement in system throughput and a \ResDDSSoATpotSUNvGPU$\times$ reduction in per-user token output latency, despite operating with \ResDDSSoAPeakRedDSProf$\times$ lower peak system performance compared to a \gls{soa} solution deployed on 96 NVIDIA H800 GPUs.
\end{itemize}

In our preliminary work \cite{zhang2025}, we introduced the FlatAttention dataflow for the original \gls{mha} architecture and provided a preliminary evaluation focusing primarily on the prefill phase.
In this work, we propose a generalized FlatAttention dataflow supporting all major attention mechanisms employed in modern \glspl{llm}.
Additionally, we extend the evaluation to both prefill and decoding.
Furthermore, we provide new insights by assessing FlatAttention in a full end-to-end inference workload.
To this end, we extend our simulation framework to model \syspl comprising multiple tile-based accelerators, interconnected via dedicated \gls{d2d} links.

\section{Background}
\label{sec:background}

\begin{figure*}[t!]
  \centering
  \includegraphics[width=1.0\linewidth]{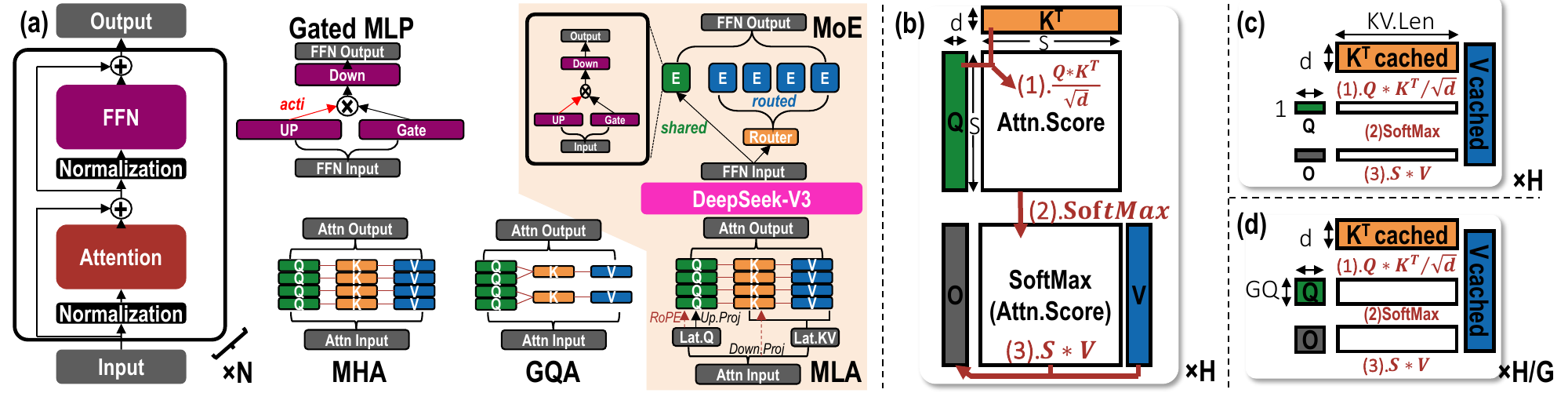}
  \begingroup
    \phantomsubcaption
    \label{fig_llm_architecture}
    \phantomsubcaption
    \label{fig_mha_prefill}
    \phantomsubcaption
    \label{fig_mha_decoding}
    \phantomsubcaption
    \label{fig_gqa_decoding}
    \phantomsubcaption
    \label{fig_mha_speculative_decoding}
  \endgroup
  \vspace{-5mm}
  \caption{(a) \gls{soa} \gls{llm} model architecture overview and schematic for (b) \gls{mha} in prefill (c) \gls{mha} in auto-regressive decoding as well as (d) \gls{gqa} in auto-regressive decoding.}
  \label{fig_background}
\end{figure*}

\subsection{\gls{soa} \glspl{llm}}
\label{subsec:bg_llm}

Many of today's \gls{soa} \glspl{llm}, such as GPT-4, LLaMA and DeepSeek-v3, employ decoder-only networks built from a stack of decoder blocks each comprising two main components: a \gls{mha} layer and a \gls{ffn}. These modules are embedded within a residual topology with normalization layers to maintain stability at scale.

\Cref{fig_llm_architecture} illustrates this generic high-level architecture, as well as common variants for the \gls{mha} and \gls{ffn} blocks, next to the \gls{soa} open source DeepSeek-v3 model architecture, a concrete implementation of this high-level architectural template.
DeepSeek-v3 employs \gls{mla} and \gls{moe} layers to implement the attention and \gls{ffn} blocks, respectively.
Beyond attention and \gls{ffn} innovations, DeepSeek-v3 employs other recent architectural improvements: \glspl{rope} to capture relative positional information directly within the attention mechanism, \gls{silu} activations, and RMSNorm for increased computational efficiency over LayerNorm.

\subsection{Attention}
\label{subsec:bg_attn}

In a standard \gls{mha} layer, an input sequence of $S$ tokens $X \in \mathbb{R}^{S \times d_{\text{model}}}$ is projected into $h$ query, key and value matrices:
\begin{equation}
Q_i = X W^Q_i,\qquad K_i = X W^K_i,\qquad V_i = X W^V_i
\end{equation}
with $W^Q_i, W^K_i, W^V_i \in \mathbb{R}^{d_{\text{model}} \times d}$ and $d$ the per-head dimensionality, to calculate:
\begin{equation}
    \begin{split}
\text{MHA}(Q, K, V) &= \text{Concat}(Attn_1, \dots, Attn_h)W^O \\
Attn_i &= \text{Softmax}(\frac{Q_i K_i^T}{\sqrt{d}})V_i
    \end{split}
\end{equation}
where $W^O \in \mathbb{R}^{d \times d_{\text{model}}}$.
The computation of one attention head $Attn_i$ is illustrated in \Cref{fig_mha_prefill}.

In autoregressive inference, an initial input sequence, or prompt, is fed to the network (\emph{prefill phase}), after which tokens are generated sequentially and appended to the input sequence (\emph{decoding phase}).
At decoding step $t$, only the query, key and value vectors of the newly generated token are computed, while key and value vectors of previous tokens are retrieved from a \emph{\gls{kv} cache} populated during prefill and prior decoding iterations.
Notably, the \gls{gemm} for both the attention score and output calculations reduce to matrix-vector multiplications (GEMV), as illustrated in \Cref{fig_mha_decoding}, for all decoding steps.
However, the full $K_i$ and $V_i$ matrices are still used as part of this calculation; thus, at every decoding step, the \gls{kv} cache grows linearly with the sequence length $S$, increasingly stressing memory bandwidth and motivating more efficient attention variants.

In \acrfull{mqa} \cite{shazeer2019fast}, adopted by LLaMA~2, key and value projections are shared by all heads, reducing the KV cache size by $h$ times:
\begin{equation}
    W_i^K = W^K,\qquad W_i^V = W^V,\qquad \forall i \in [1, h]
\end{equation}
\acrfull{gqa} \cite{ainslie2023gqa}, adopted by LLaMA~3, generalizes \gls{mqa} by sharing keys and values across groups of heads, providing a tunable trade-off between cache size, compute efficiency, and model accuracy:
\begin{equation}
    W_i^K = W^K_{g(i)},\qquad W_i^V = W^V_{g(i)}
\end{equation}
As $G$ heads in a group share the same $K_{g(i)}$, $V_{g(i)}$ matrices, their respective queries can be concatenated to turn the attention score and output computations back into \glspl{gemm}, as shown if \Cref{fig_gqa_decoding}.

Instead of sharing key and value projections across heads, \acrfull{mla}, introduced in DeepSeek-v2 \cite{deepseekv2}, compresses each head's key and value vectors by mapping them into a compact latent space, where the shared low-rank down-projections for the layers are:
\begin{equation}
    c^{KV} = XW^{DKV}, \qquad c^{Q} = XW^{DQ}
\end{equation}
The attention projections are then decompressed from the latent space:
\begin{equation}
Q_i = c^{Q} W^{UQ}_i, \qquad K_i = c^{KV} W^{UK}_i,\qquad V_i = c^{KV} W^{UV}_i
\end{equation}
with $W^{DKV} \in \mathbb{R}^{d_{\text{model}} \times d_c}$ a shared down-projection matrix for both keys and values, and $W^{UK}_i, W^{UV}_i \in \mathbb{R}^{d_c \times d}$ unique up-projection matrices.
Only the compressed keys and values need to be cached, optimizing the \gls{kv} cache size.

Apart from auto-regressive decoding, speculative decoding has emerged as an effective technique for reducing inference latency in large language models.
Speculative decoding leverages a lightweight draft model to generate multiple candidate tokens ahead of time, which are then verified in parallel by a larger target model.
Some models further support \emph{self-speculative} decoding, where the target model itself produces draft tokens.
Similar to \gls{gqa}, also speculative decoding can help restore GEMV operations into \gls{gemm} operations.

\subsection{\acrfull{ffn}}
\label{subsec:bg_ffn}

In the \gls{ffn} block, modern LLMs commonly adopt a gated \gls{mlp} structure (\Cref{fig_llm_architecture}), which consists of an \emph{up-projection} that expands the hidden dimension, a \emph{gate} projection that modulates activation through element-wise gating, and a \emph{down-projection} that projects the activations back to the original hidden dimension.
This gated design improves expressiveness and training stability while maintaining computational efficiency.

To improve scaling efficiency, DeepSeek-v3 replaces the dense \gls{ffn} with a \acrfull{moe} architecture.
In a \gls{moe} layer, a learned gating network routes each token to a small subset of experts---specialized \glspl{ffn}---instead of activating all of them, as shown in \Cref{fig_llm_architecture}. If only $k$ out of $E$ experts are selected per token, the computational cost becomes proportional to $k$ rather than $E$, while $E$ and the total parameter count can be increased freely. This enables higher model capacity without a proportional increase in inference cost.

\subsection{Tile-Based Many-\gls{pe} Architecture}
\label{subsec:bg_arch}

As illustrated in \Cref{fig_arch}, the fundamental building block of the many-\gls{pe} architecture is the ``tile''.
Each tile comprises \glspl{pe}, local memory (L1), a \gls{dma} engine, and local interconnects.
There are three main types of \glspl{pe}: scalar cores, vector engines, and matrix engines.
Scalar cores mainly handle dataflow control tasks, whereas heavy computational tasks are offloaded to the vector and matrix engines based on the computation type.
The local L1 memory is implemented as a software-managed scratchpad memory for increased area efficiency; all \glspl{pe} within a tile can directly access the local L1 memory via the local interconnect.
The \gls{dma} engine in each tile is responsible for bulk data movement in and out of the local L1 memory. 

The tile-based many-\gls{pe} system uses an on-chip 2D-mesh \gls{noc} to connect tiles.
Off-chip memory, such as \gls{hbm}, is located at the boundary of the mesh \gls{noc}, interfaced through the respective memory controllers.
To improve the efficiency of data exchange among tiles, fabric-level acceleration of collective communication primitives has been proposed \cite{jerger2008,krishna2011,ma2012,colagrande2026}.
Using a row-wise multicast as an illustrative example in \Cref{fig_coll_demo}, hardware-supported collective primitives perform fine-grained, flit-level data replication within \gls{noc} routers along the multicast path, enabling all destinations to receive data with low latency.
This approach is significantly more efficient than software-managed collectives, which require multiple stages of point-to-point transfers.

The tile-based accelerator designs can be replicated across multiple dies and integrated into a wafer-scale system.
Wafer-scale integration can be realized using either monolithic fabrication on a single wafer\cite{lie2022cerebras,zhang2024benchmarking} or chiplet-based integration\cite{mishty2024chiplet,xu2025wsc}. In the latter approach, compute, memory and IO dies are fabricated separately and integrated on a wafer-scale silicon interposer using Chip-on-Wafer-on-Substrate (CoWoS) packaging technology\cite{cowos,xu2025wsc}. The dies are electrically interconnected through through-silicon vias (TSVs) and redistribution layers (RDLs) in the interposer, enabling greater heterogeneity and design flexibility, as illustrated in \Cref{fig_multi_die_arch}.
In addition to compute dies hosting tile-based accelerators, the system integrates dedicated HBM memory dies and router dies to support wafer-scale \gls{d2d} interconnect.
In this work, we adopt a conventional 2D mesh as the baseline interconnect topology.

\section{Dataflow Implementation}
\label{sec:impl}

\subsection{Motivation}
\label{subsec:motivation}

\begin{figure}[t!]
  \centering
  \includegraphics[width=1\linewidth, trim={2.5cm 8.7cm 4cm 8.25cm}, clip]{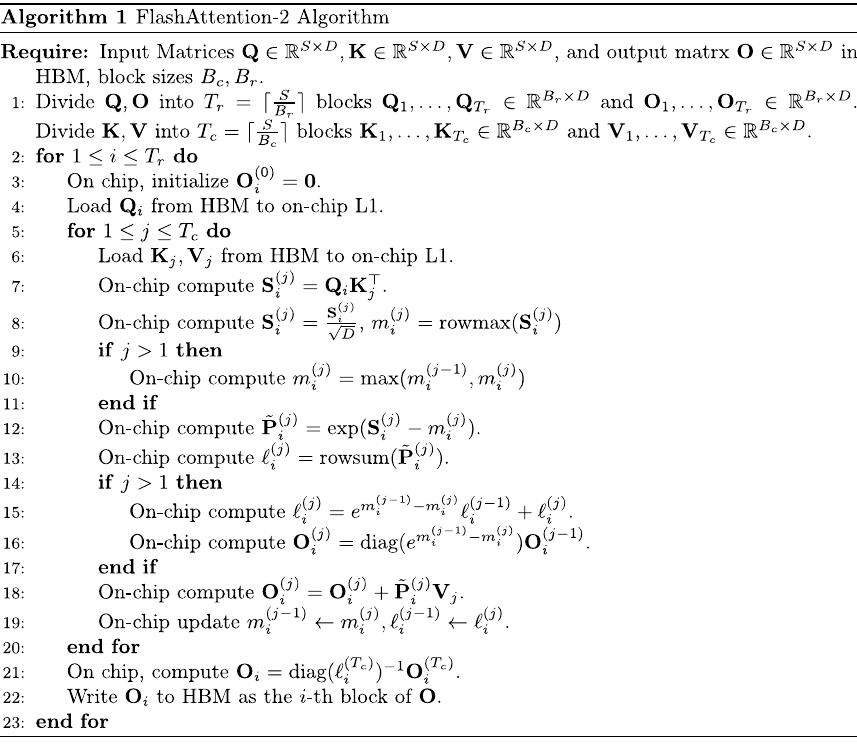}
  \vspace{-10mm}
  \label{alg_flashattention_2}
\end{figure}

\begin{figure*}[t!]
  \centering
  \includegraphics[width=1.0\linewidth, trim={0.7cm 0.2cm 0.7cm 0.2cm}, clip]{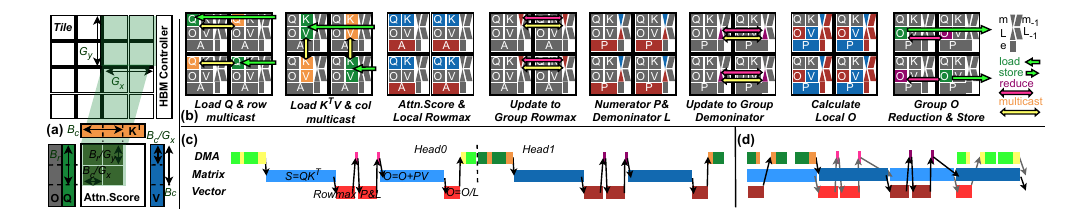}
  \begingroup
    \phantomsubcaption
    \label{fig_flat_mapping}
    \phantomsubcaption
    \label{fig_flat_dataflow}
    \phantomsubcaption
    \label{fig_flat_naive_schedule}
    \phantomsubcaption
    \label{fig_flat_async_schedule}
  \endgroup
  \vspace{-5mm}
  \caption{(a) Parametric definition of FlatAttention. (b) Detailed FlatAttention dataflow. (c) Naive FlatAttention schedule. (d) Optimized asynchronous FlatAttention schedule.}
  \label{fig_flat}
\end{figure*}

\begin{figure}[t!]
  \centering
  \includegraphics[width=1\linewidth,trim={2.5cm 6cm 4cm 5.12cm}, clip]{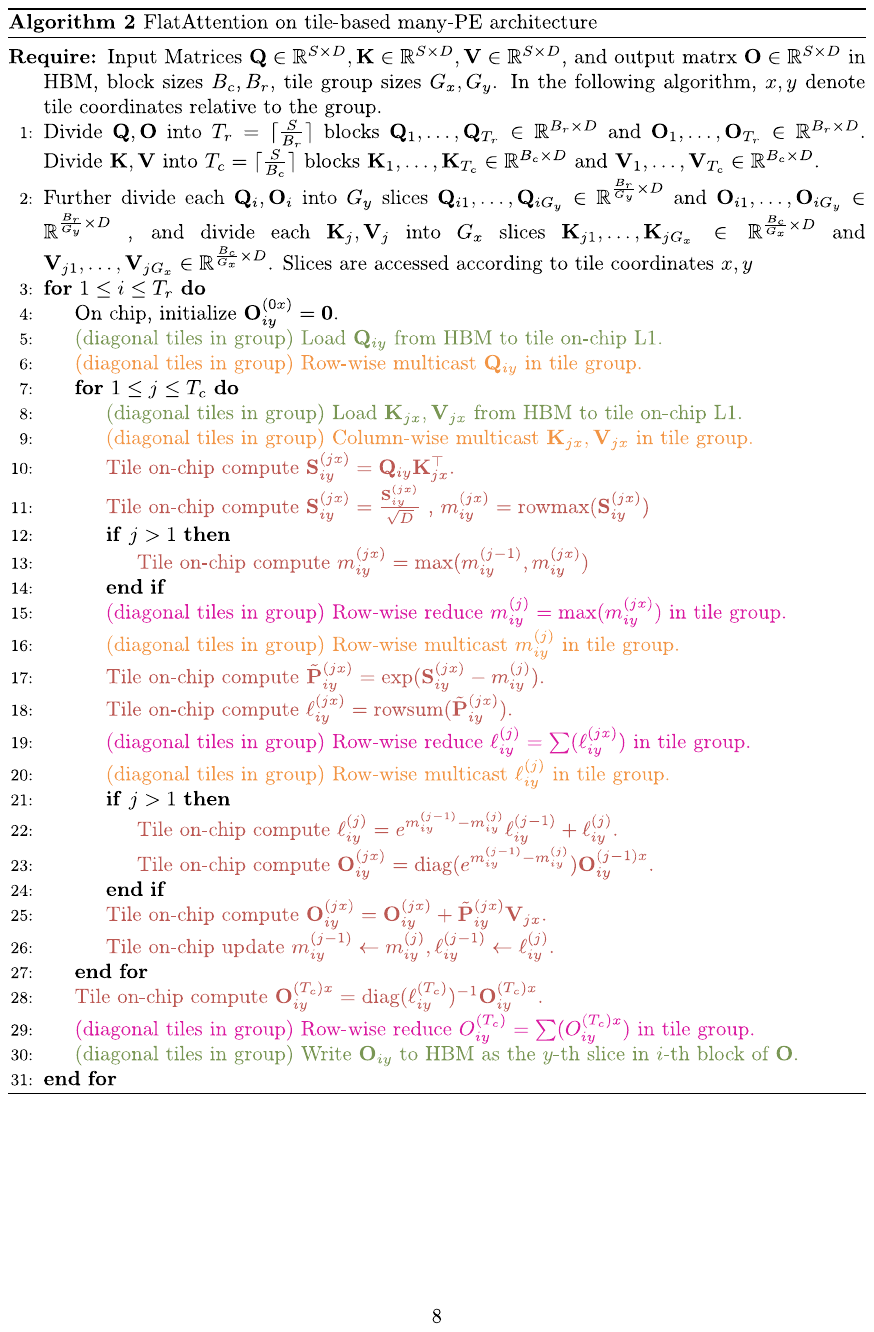}
  \vspace{-10mm}
  \label{alg:flat}
\end{figure}

FlashAttention is the first widely adopted \gls{mha} dataflow implementation for \glspl{gpu} that efficiently fuses matrix multiplication and softmax operations within each attention head, maximizing data reuse in the shared memory of each \gls{sm}.
The FlashAttention-2 algorithm is illustrated in Algorithm 1.
In this approach, the \gls{mha} workload is partitioned across the batch and head dimensions over the \glspl{sm}, and, when the batch size or number of heads is insufficient, additionally along the output sequence length dimension.

This dataflow and mapping strategy can also be applied to tile-based many-\gls{pe} accelerators, since an \gls{sm} in a GPU is analogous to a compute tile in such architectures. With this mapping, each compute tile operates on distinct data, as different heads do not share or reuse data, allowing them to access \gls{hbm} independently. This eliminates the need for inter-tile communication, but also prevents any data reuse across tiles.
In prefill mode of \gls{mha}, with sequence length $S$, head dimension $D$, number of heads $H$, batch size $B$ and block size $M \coloneqq B_r = B_c$, the FlashAttention dataflow results in an HBM I/O complexity of:
\vspace{-1mm}
\[
\text{IO}
= B \cdot H \cdot N_{\text{outer}}
\cdot \Bigl(
Q_{\text{block}} + O_{\text{block}}
+ N_{\text{inner}} \cdot (K_{\text{block}} + V_{\text{block}})
\Bigr).
\]

With the number of outer and inner loop iterations
$N_{\text{outer}} = N_{\text{inner}} = \frac{S}{M}$,
and block sizes
$Q_{\text{block}} = O_{\text{block}} = K_{\text{block}} = V_{\text{block}} = D \cdot M$,
the expression simplifies to:
\vspace{-2mm}
\[
\text{IO}
= 2 \cdot B \cdot H \cdot D \cdot S
\cdot \left( 1 + \frac{S}{M} \right).
\]

While all other parameters are fixed by the computation, the block size parameter $M$ can be increased to reduce I/O complexity. Intuitively, larger blocks improve data reuse within a tile’s L1 memory.
However, $M$ is constrained by the capacity of the L1 memory on a single tile, which must be able to simultaneously store the tensors $Q_i, K^T_j, V_j,$ and $O_i$ at any point in time.

To further reduce off-chip I/O accesses, we propose FlatAttention, a dataflow that fundamentally redefines how \gls{mha} is parallelized on tile-based architectures. FlatAttention treats multiple tiles as a unified compute unit, enabling them to collaboratively process an \gls{mha} block—defined above—of substantially larger size by leveraging the aggregate L1 memory of the tile group to collectively store the block.

Given that a single tile can host a block of size $M \coloneqq B_r = B_c$, grouping $\textcolor{red}{N} \times \textcolor{red}{N}$ tiles allows the group to collectively host a larger block $(\textcolor{red}{N} B_r, \textcolor{red}{N} B_c)$. As a result, the I/O complexity of \gls{mha} becomes:

\[
\text{IO} = 2 \cdot B \cdot H \cdot D \cdot S \cdot \left( 1 + \frac{ S}{\textcolor{red}{N} \cdot M} \right).
\]

For example, when $S = 4096$, $M = 128$, and $N = 8$, this yields a theoretical $6.6\times$ reduction in HBM accesses compared to the FlashAttention dataflow.
However, the reduction in HBM traffic achieved by our proposed FlatAttention dataflow comes at the cost of introducing on-chip inter-tile communication, which is required to collectively process a large \gls{mha} block within the tile group.
\emph{Efficiently handling on-chip inter-tile communication therefore becomes crucial to realizing the benefits of reduced off-chip communication}.

\subsection{Detailed FlatAttention Dataflow}
\label{subsec:flat}

The FlatAttention dataflow is depicted in \Cref{fig_flat}.
We refer to a set of tiles collectively processing a block, as previously introduced, as a \textit{group}, illustrated in \Cref{fig_flat_mapping}.
We define the shape of the group as $G_x \times G_y$. FlatAttention applies the same tiling and mapping scheme to \emph{groups} as FlashAttention applies to \emph{tiles}, but it introduces a secondary level of blocking within each group. This secondary blocking divides the $\{B_c, B_r\}$ block dimensions into smaller \textit{slices} based on the group shape $\{G_x, G_y\}$, resulting in $\{\frac{B_c}{G_x}, \frac{B_r}{G_y}\}$ slice sizes for every tile.

Algorithm 2 and \Cref{fig_flat_dataflow} outline the FlatAttention dataflow.
At a high level, the algorithm is conceptually similar to FlashAttention (\Cref{alg_flashattention_2}): different groups process distinct data, so no communication between groups is required. However, distributing the computation of an \gls{mha} block to tiles in a group introduces distinct data movement patterns within the group:

\begin{itemize}
    \item \textbf{Loading and Multicasting}: 
    Only diagonal tiles of the group load $Q$ slices from HBM (line \tikz[baseline, yshift=4pt]{\node[shape=circle, draw, fill=color_ls, text=white, inner sep=0.5pt, font=\small] {5};}), followed by multicasting $Q$ slices row-wise \tikz[baseline, yshift=4pt]{\node[shape=circle, draw, fill=color_mc, text=white, inner sep=0.5pt, font=\small] {6};} to the other tiles in the group.
    When entering the inner loop, the diagonal tiles load $K$ and $V$ slices from HBM \tikz[baseline, yshift=4pt]{\node[shape=circle, draw, fill=color_ls, text=white, inner sep=0.5pt, font=\small] {8};} and multicast them column-wise \tikz[baseline, yshift=4pt]{\node[shape=circle, draw, fill=color_mc, text=white, inner sep=0.5pt, font=\small] {9};}.

    \item \textbf{Computing Attention ($Q \cdot K^T$) and Rowmax}:
    Each tile computes a segment of the attention score matrix \tikz[baseline, yshift=4pt]{\node[shape=circle, draw, fill=color_cp, text=white, inner sep=0.5pt, font=\footnotesize] {10};}. During the computation of row-wise maxima for Softmax, tiles compute partial row maxima locally\tikz[baseline, yshift=4pt]{\node[shape=circle, draw, fill=color_cp, text=white, inner sep=0.5pt, font=\footnotesize] {11};} updated with the tracking maxima\tikz[baseline, yshift=4pt]{\node[shape=circle, draw, fill=color_cp, text=white, inner sep=0.5pt, font=\footnotesize] {13};}, followed by a row-wise reduction within the group to calculate the global row maxima\tikz[baseline, yshift=4pt]{\node[shape=circle, draw, fill=color_rd, text=white, inner sep=0.5pt, font=\footnotesize] {15};}. The results are then multicast row-wise to ensure that each tile holds the global row maxima \tikz[baseline, yshift=4pt]{\node[shape=circle, draw, fill=color_mc, text=white, inner sep=0.5pt, font=\footnotesize] {16};}.

    \item \textbf{Softmax Denominator}:
    After computing the partial Softmax denominator locally with global row maxima\tikz[baseline, yshift=4pt]{\node[shape=circle, draw, fill=color_cp, text=white, inner sep=0.5pt, font=\footnotesize] {17};}\tikz[baseline, yshift=4pt]{\node[shape=circle, draw, fill=color_cp, text=white, inner sep=0.5pt, font=\footnotesize] {18};}, the same reduction \tikz[baseline, yshift=4pt]{\node[shape=circle, draw, fill=color_rd, text=white, inner sep=0.5pt, font=\footnotesize] {19};} and multicast \tikz[baseline, yshift=4pt]{\node[shape=circle, draw, fill=color_mc, text=white, inner sep=0.5pt, font=\footnotesize] {20};} procedure applies to computing the global denominator, which is then updated with the tracking maxima and denominator \tikz[baseline, yshift=4pt]{\node[shape=circle, draw, fill=color_cp, text=white, inner sep=0.5pt, font=\footnotesize] {22};}.

    \item \textbf{Output Matrix ($O$)}:
    Each tile updates local $O$ slices and tracking statistics in the inner loop, and computes partial results for $O$ slices on exit \tikz[baseline, yshift=4pt]{\node[shape=circle, draw, fill=color_cp, text=white, inner sep=0.5pt, font=\footnotesize] {23};}-\tikz[baseline, yshift=4pt]{\node[shape=circle, draw, fill=color_cp, text=white, inner sep=0.5pt, font=\footnotesize] {28};}. FlatAttention then performs a row-wise reduction of $O$ slices \tikz[baseline, yshift=4pt]{\node[shape=circle, draw, fill=color_rd, text=white, inner sep=0.5pt, font=\footnotesize] {29};} followed by storing $O$ slices in \gls{hbm} \tikz[baseline, yshift=4pt]{\node[shape=circle, draw, fill=color_ls, text=white, inner sep=0.5pt, font=\footnotesize] {30};} only from diagonal tiles.
\end{itemize}

These communication requirements are a direct result of FlatAttention's parallelization scheme, which enables minimizing costly \textit{global} off-chip I/O by exploiting on-chip data reuse across tiles through \textit{local} on-chip communication.
This trade-off of global for local requirements enables FlatAttention to achieve better scalability and performance compared to FlashAttention methods for tile-based many-PE architectures, as long as local on-chip communication is efficiently handled, as will be discussed in \Cref{subsec:exp1}.

\subsection{Asynchronous FlatAttention}
\label{subsec:opti}

In the na\"ive version of FlatAttention (Algorithm 2), data movement and Softmax-related computations still account for a significant portion of the runtime, as illustrated in \Cref{fig_flat_naive_schedule}.
This reduces the overall utilization, as the system's peak performance is primarily determined by the matrix engine, which has much higher computational power compared to the vector engine. To further improve utilization, we propose leveraging the asynchronous nature of \gls{dma}, vector and matrix engine invocations to overlap the runtime of data movement and Softmax operations with matrix multiplications.

The optimized dataflow schedules the computation of two heads concurrently on each group. While the matrix engine processes matrix multiplications for one head, the \gls{dma} and vector engine perform data movement and Softmax operations for the other\footnote{The same optimization can be applied with two output row blocks $O_i$ instead of two heads, reducing memory requirements as the $K^T_j$ and $V_j$ blocks are shared. To simplify the evaluation, where sufficient row blocks are not available, we adopt the presented implementation.}. \Cref{fig_flat_async_schedule} demonstrates this optimization, showcasing how it can ensure that matrix engine remains nearly always active, provided that the runtime of the matrix multiplication overlaps completely with data movement and Softmax operations. 
Notably, FlashAttention-3 employs a similar technique to improve upon FlashAttention-2, though the exact implementation varies due to architectural differences.

\subsection{Generalization to Decode Stage and Attention Variants}
\label{subsec:generalization}

Beyond the prefill-phase \gls{mha} dataflow, FlatAttention can be generalized to the decode phase, as well as to attention variants such as \gls{gqa} and \gls{mla}. In the auto-regressive decode phase, the sequence length of $Q$ is no longer equal to that of $KV$.
Instead, the sequence length of $Q$ becomes $1$, while the sequence length of $KV$ corresponds to the KV-cache length, as illustrated in \Cref{fig_mha_decoding}.
To apply the FlatAttention dataflow to \gls{mha} in the auto-regressive decode phase, the tile group can be configured to span a single row, processing a block size of $B_r = 1$ for $Q$.
Meanwhile, the block size $B_c$ can be increased to maximize reuse of the KV cache by leveraging the aggregated L1 memory capacity of the tile group.

For \gls{mha} in speculative decoding, the sequence length of $Q$ increases from $1$ to the speculative length $sp$ as shown in \Cref{fig_mha_speculative_decoding}, with causal mask applied. 
Depending on the sequence lengths of $Q$ and $KV$, an appropriate tile-grouping and blocking scheme can be applied for FlatAttention, which will be discussed in \Cref{subsec:exp2}.

In the case of \gls{gqa}, multiple query heads share a single KV cache. As a result, the grouped queries can be interpreted as forming an effective query sequence of increased length $1*G$, while attending to the same KV data. The decode operation can thus be viewed as executing  $\frac{H}{G}$ attention computations in parallel over a longer query sequence, as illustrated in \Cref{fig_gqa_decoding}.

To improve decoding efficiency, \gls{mla} is used together with a weight-absorption trick that converts it into \gls{mqa} mode. The key observation is that the problematic up-projection matrices $W^{UK}_i$ and $W^{UV}_i$ always appear inside matrix products where the multiplications can be rearranged by associativity. For example, to absorb $W^{UK}_i$ into the query side, consider the attention score term $Q_i K_i^{\top}$ for head $i$ and rewrite it in terms of the up-projection matrices as follows:
\begin{equation}
    Q_iK_i^T = (c^QW^{UQ}_i)(c^{KV}W^{UK}_i)^\top = c^QW^{UQ}_iW^{UK\top}_ic^{KV\top}
\end{equation}
Where the $W^{UK}_i$ can be absorbed into $W^{UQ}_i$ as:
\begin{equation}
    W^{UQK}_i = W^{UQ}_iW^{UK\top}_i
\end{equation}
By further transformation, the \gls{mla} formulation can be viewed as \gls{mqa}, where $c^{Q}$ is projected with $W^{UQK}_i$ to produce all query heads that share the same key–value representations derived from $c^{KV}$.
In this case, the generalization of the FlatAttention dataflow is similar to \gls{mqa}, where the queries share a grouped sequence length across all heads. The detailed formulations of the weight-absorption transformation and the corresponding FlatAttention generalization for \gls{mla} in the DeepSeek-v3 models are presented in Appendix A.

We observe that these modern attention variants can all be transformed into a unified multi-head attention (MHA) formulation. 
They primarily differ in the shape of the attention score matrices and the number of attention heads across variants. 
A key challenge for the performance of FlatAttention lies in mapping irregularly shaped attention score matrices and determining the appropriate scale of tile groups. 
We elaborate on this in detail in \Cref{subsec:exp2}.

\subsection{DeepSeek-v3 Decoder and \gls{gemm} Dataflow}
\label{subsec:ds_decode}

\begin{figure*}[t!]
    \centering
    \includegraphics[width=1.0\linewidth, trim={0.7cm 0.2cm 0.7cm 0.2cm}, clip]{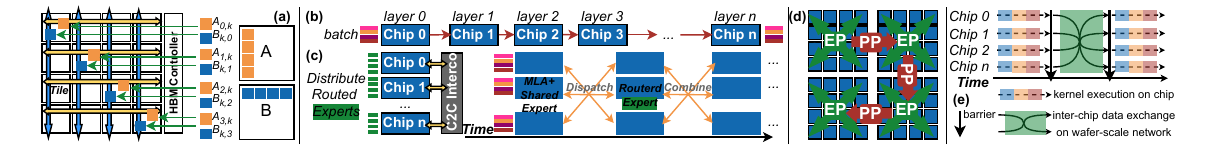}
    \begingroup
        \phantomsubcaption
        \label{fig_summa_dataflow}
        \phantomsubcaption
        \label{fig_pipeline_parallelism}
        \phantomsubcaption
        \label{fig_expert_parallelism}
        \phantomsubcaption
        \label{fig_hybrid_parallelism}
        \phantomsubcaption
        \label{fig_wafer_system_time_model}
    \endgroup
    \vspace{-3mm}
    \caption{(a) General SUMMA dataflow on tile-based accelerators for \gls{gemm}. DeepSeek-v3 workload distribution in (b) pipeline parallelism, (c) full expert parallelism, and (d) EP-PP hybrid parallelism. (e) Wafer-scale multi-chip system execution mode.}
    \label{fig_dsv3_flow}
\end{figure*}

To evaluate the impact of the FlatAttention dataflow speedup on end-to-end \gls{llm} inference performance, we adopt the DeepSeek-v3-671B\cite{deepseekv3} decoder block as a case-study application.
DeepSeek-v3-671B enables \gls{mtp} by default, which performs self-contained speculative decoding by predicting one next token per decoding iteration. This corresponds to a speculative length of 2, with an acceptance rate of 0.7 for the predicted token.
A DeepSeek-v3-671B layer is composed of a sequence of computational kernels, including normalization, projections, RoPE, the core MLA mechanism, among others.
The detailed kernel flow for a DeepSeek-v3-671B decoding layer is presented in Appendix B.
In our evaluation, we execute kernels sequentially, running one kernel at a time on the \arch accelerator.

Among all kernels, \gls{gemm} appears in every projection kernel. We implement \gls{gemm} on a tile-based many-\gls{pe} accelerator using the SUMMA\cite{summa} algorithm, as illustrated in \Cref{fig_summa_dataflow}.
In the SUMMA dataflow, each tile stationarily computes an output block $C_{i,j}$. At every iteration over the $K$ dimension, a column of blocks $A_{i,k}$ is multicasted row-wise, while a row of blocks $B_{k,j}$ is multicasted column-wise across tiles.
If the column-wise multicasts were initiated from tiles in the same row, the read requests to the \gls{hbm} would conflict on the same \gls{noc} links.
Therefore, we assign the task of fetching data from \gls{hbm} and performing subsequent row- and column-wise multicasts to the tiles along the diagonal.

\subsection{DeepSeek-v3 Decoder Mapping on Wafer-Scale Multi-Die System}
\label{subsec:dsv3sys}

As the parameter size reaches 671B, a single accelerator chip with \glspl{hbm} typically provides limited memory capacity (e.g. 96-144 GiB in GH200), which is insufficient to store all parameters at the default FP8 precision.
Consequently, distributing the DeepSeek-v3 decoder workload across multiple chips is necessary, and various parallelism strategies must be considered.

Two key metrics are commonly used to evaluate decoder performance: overall throughput (tokens/s) and \gls{tpot}. 
The \gls{tpot} metric captures per-user output token latency and is typically targeted to be within 50\,ms. Many parallelism paradigms have been proposed to increase throughput while simultaneously reducing \gls{tpot} in \gls{llm} decoder systems.

In this study, we model and discuss \gls{pp}, \gls{ep}, and hybrid approaches combining the two. 
In a typical \gls{pp} organization, shown in \Cref{fig_dsv3_flow}e, each chip is assigned one or more decoder layers, and \gls{c2c} communication only transfers intermediate activations between layers. 
In contrast, full \gls{ep} organization is fundamentally different: each chip executes the complete decoder stack, except for the routed experts in the \gls{moe}, whose weights are distributed across chips.
During decoding, different users’ token batches are processed in parallel across all chips for the \gls{mla} and shared experts of the \gls{moe}, followed by \gls{c2c} communication to dispatch tokens to the chips hosting the selected routed experts. After expert computation, \gls{c2c} communication is used again to gather results.

Hybrid parallelism combines \gls{pp} and \gls{ep}. For example, an EP16–PP4 configuration applies expert parallelism within groups of 16 chips, while pipeline parallelism is used across 4 groups.

\section{Modeling and Analysis Methodology}
\label{sec:meth}

\begin{figure*}[t]
    \centering
    \begin{subfigure}{0.8\columnwidth}
        \centering
        \includegraphics[width=\textwidth]{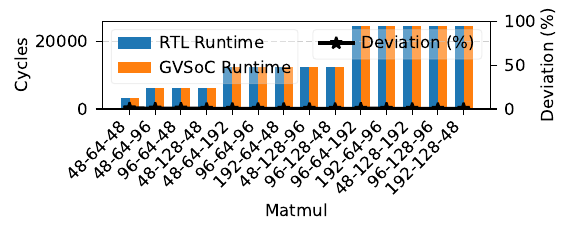}
        \put(-125,6.3){\small (a)}
        \captionsetup{labelformat=empty}
        \caption{}
        \vspace{-1.3em}
        \label{plot_RedMule_Calibration}
    \end{subfigure}
    \begin{subfigure}{0.6\columnwidth}
        \centering
        \includegraphics[width=\textwidth]{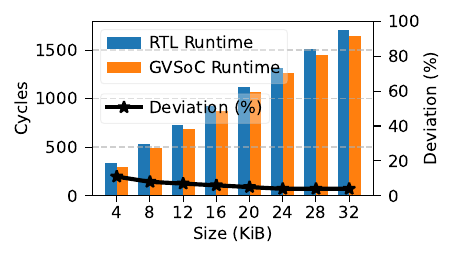}
        \put(-105,6.3){\small (b)}
        \captionsetup{labelformat=empty}
        \caption{}
        \vspace{-1.3em}
        \label{plot_NoC_Seq_Calibration}
    \end{subfigure}
    \begin{subfigure}{0.6\columnwidth}
        \centering
        \includegraphics[width=\textwidth]{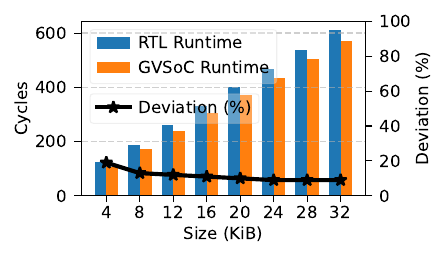}
        \put(-105,6.3){\small (c)}
        \captionsetup{labelformat=empty}
        \caption{}
        \vspace{-1.3em}
        \label{plot_NoC_HW_Calibration}
    \end{subfigure}
    \vspace{-0.6em}
    \caption{GVSoC–RTL calibration for (a) RedMulE and NoC under (b) \textit{SW.Seq} row-wise multicast and (c) \textit{HW} row-wise reduction collective patterns.}
    \label{plot_calibration}
    \vspace{-0.5em}
\end{figure*}

We developed a modeling and simulation framework for tile-based many-PE accelerators on the GVSoC event-based simulator \cite{gvsoc} for functional and performance simulation.
GVSoC is open source and released with models for the single-issue RISC-V core, the Spatz\cite{perotti2025spatz} vector engine supporting the \gls{rvv} extension, the DMA engine, and tile-local L1 memory and interconnect.
To extend these capabilities, we developed and calibrated new models for the RedMulE\cite{tortorella2023RedMulE} matrix engine and the FlooNoC fabric according to their open-source RTL implementations.
Colagrande et al. propose a lightweight extension to the FlooNoC RTL to support collective communication primitives \cite{colagrande2026}. 
We extend GVSoC's \gls{noc} model with hardware support for the collective communication primitives described in \Cref{subsec:bg_arch}, modeled after the previous work.
We further extend Spatz with a custom \gls{rvv} instruction to support exponential operations, along with a dedicated exponential unit integrated into the FPU, modeled and calibrated after the implementation described in~\cite{pace}.
Furthermore, we integrated the performance-accurate DRAMSys \cite{jung2015dramsys} simulator into GVSoC for HBM modeling.

We calibrate the GVSoC model of RedMulE using the default $12 \times 4$ \gls{ce} array configuration across a range of matrix multiplication shapes. 
As shown in \Cref{plot_RedMule_Calibration}, the average cycle deviation is 0.17\%.
We further calibrate the GVSoC NoC model using the default $4 \times 4$ mesh configuration under two collective communication patterns: \textit{SW.Seq} row-wise multicast and \textit{HW} row-wise reduction. 
The results, shown in \Cref{plot_NoC_Seq_Calibration} and \Cref{plot_NoC_HW_Calibration}, exhibit average cycle deviations of 6\% and 12\%, respectively.

Using these building blocks, we constructed the \textbf{SoftHier} model and analysis framework on the GVSoC platform: a flexible, parameterizable tile-based many-PE accelerator simulator.
The framework is configurable using architecture configuration files, enabling the instantiation of specific accelerator designs, e.g. to explore different numbers of \glspl{ce} in the RedMulE units.
In the SoftHier framework, we implemented the kernel dataflows in C and compiled them using the \texttt{GNU RISC-V GCC} compiler with \texttt{-O3} optimization.

To evaluate wafer-scale system performance, we employ a naive parallel execution model, illustrated in \Cref{fig_wafer_system_time_model}, in which kernel execution on individual chips and chip-to-chip communication are fully separated by synchronization barriers. 
Under this execution model, system performance can be estimated by combining kernel runtimes measured on individual accelerators modeled in SoftHier, with chip-to-chip communication latency obtained from a separate \gls{c2c} model in the GVSoC simulator.
The \gls{c2c} model abstracts each chip as a traffic generator while explicitly modeling the \gls{d2d} links to provide a high-level representation of existing interconnect protocols (e.g., NVLink and CXL). 
This abstraction supports credit-based flow control with configurable latency and bandwidth, and the modeled \gls{d2d} routers handle packet-level congestion.

\section{Experimental Results}
\label{sec:expe}

\begin{figure}[t]
    \centering
    \begin{subfigure}{0.48\columnwidth}
        \centering
        \includegraphics[width=\textwidth]{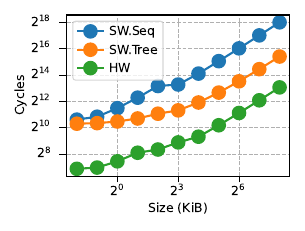}
        \put(-90,7.3){\small (a)}
        \captionsetup{labelformat=empty}
        \caption{}
        \vspace{-1.5em}
        \label{fig_Multicast_Comparision}
    \end{subfigure}
    \begin{subfigure}{0.48\columnwidth}
        \centering
        \includegraphics[width=\textwidth]{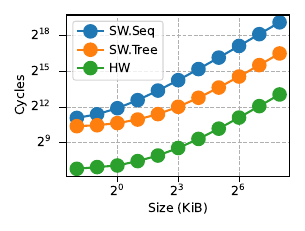}
        \put(-90,7.3){\small (b)}
        \captionsetup{labelformat=empty}
        \caption{}
        \vspace{-1.5em}
        \label{fig_RedSum_Comparision}
    \end{subfigure}
    \vspace{-0.6em}
    \caption{Latency comparison of software-based and fabric-accelerated collective primitives: (a) row-wise multicast; (b) row-wise sum reduction on accelerator with 32$\times$32 tiles.}
    \label{plot_noc_characteristic}
    \vspace{-0.5em}
\end{figure}

\begin{table}[ht]
    \centering
    \caption{System Specifications}
    \label{tab:system_specs1}
    \renewcommand{\arraystretch}{1.2} 
    \setlength{\tabcolsep}{8pt} 
    \begin{tabular}{@{}l@{ }l@{}}
        \toprule
        \textbf{Chip} & 32×32 Tiles, 1024-bit NoC link width, run at 965MHz \\ 
        \midrule
        \textbf{HBM} & HBM4 stack with 32 channels on south edge\\ 
        \textbf{Tile} & \textbf{RedMulE Matrix.Eng}: 32×16 \,CEs, 1024 FLOP/cyc@FP16 \\ 
        & \textbf{Spatz Vector.Eng}: 4 Spatz, each with 32 FLOP/cyc@FP16 \\ 
        & \textbf{Local Memory}: 384 KiB, 512 Byte/cyc \\ 
        \textbf{Summary} & 988\,TFLOPS@FP16 Peak\,Performance\, \\
        & 2\,TB/s \,Peak\,HBM\,Bandwidth \\ 
        \bottomrule
    \end{tabular}
    \vspace{-3mm}
\end{table}

\subsection{FlashAttention vs. FlatAttention}
\label{subsec:exp1}

We first characterize the performance of fabric collective primitives on a large-scale \arch accelerator, as specified in \Cref{tab:system_specs1}. The evaluated system consists of a $32 \times 32$ mesh of tiles and delivers the same peak FP16/BF16 performance as the NVIDIA GH200 GPU.

We study two representative collective communication patterns used in the proposed FlatAttention dataflow, namely row-wise multicast and row-wise sum reduction. For each pattern, we compare the performance of fabric-supported hardware collectives (\textit{HW}) against two software-based collective implementations (\textit{SW.Seq} and \textit{SW.Tree}).

As illustrated in \Cref{fig_coll_demo}, \textit{SW.Seq} represents the most naive software implementation of row-wise multicast, in which the source tile issues sequential unicast transfers to all destination tiles. In contrast, \textit{SW.Tree} employs a logarithmic tree-based multicast scheme, which significantly reduces the number of communication steps required. However, this approach requires synchronization between consecutive steps.

As shown by the results for row-wise multicast and row-wise sum reduction in \Cref{fig_Multicast_Comparision,fig_RedSum_Comparision}, respectively, fabric-supported collectives achieve substantial performance improvements as the transfer size increases. On a $32 \times 32$ mesh, hardware multicast delivers speedups of $\ResAMulticastSuSwtree\times$ and $\ResAMulticastSuSwseq\times$ over \textit{SW.Tree} and \textit{SW.Seq}, respectively, while hardware reductions deliver speedups of $\ResARedsumSuSwtree\times$ and $\ResARedsumSuSwseq\times$ over the software-based implementations.

We next compare different prefill-phase \gls{mha} implementations across multiple layer configurations. Specifically, we evaluate FlashAttention-2 (\textit{FA-2}) and FlashAttention-3 (\textit{FA-3}) on \arch accelerators, where \textit{FA-3} incorporates a scheduling strategy similar to the optimization described in \Cref{subsec:opti}. For FlatAttention, we configure a single group spanning the entire system, i.e., $G_x = G_y = 32$.

We evaluate several FlatAttention variants: FlatAttention with sequential software collectives (\textit{FlatSC}), FlatAttention with tree-based software collectives (\textit{FlatTC}), FlatAttention with hardware-supported \gls{noc} collectives (\textit{FlatHC}), and the optimized FlatAttention dataflow (\textit{FlatAync}) introduced in \Cref{subsec:opti}, which also leverages \gls{noc} collective primitives. Our evaluation spans multiple \gls{mha} layers, varying the sequence length $S \in \{1024, 2048, 4096\}$ and head dimension $D \in \{64, 128\}$, while fixing the batch size to $B = 2$ and the number of heads to $H = 32$.

\begin{figure}[t!]
  \centering
  \includegraphics[width=\columnwidth]{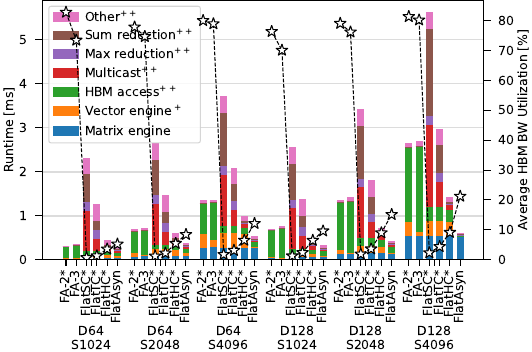}
  \caption{Runtime breakdown (bars) and average HBM BW utilization (star markers) for different \gls{mha} implementations and layer sizes. $^+$Runtime not overlapped with matrix engine. $^{++}$Runtime not overlapped with either vector or matrix engine. *Implementations without double buffering.}
  \label{plot1}
\end{figure}

\Cref{plot1} reports the runtime breakdown and average HBM bandwidth utilization. On the evaluated \arch system, FlashAttention exhibits a strongly memory-bound execution profile, with average HBM bandwidth utilization reaching up to \ResAFlashBW\%. HBM accesses dominate the overall runtime, significantly constraining effective compute utilization. Although \textit{FA-3} adopts an optimized dataflow that overlaps matrix multiplication and Softmax computation, the already saturated HBM bandwidth leaves little headroom for additional performance gains. Moreover, the more sophisticated scheduling employed by \textit{FA-3} introduces non-negligible control overhead, further diminishing its potential benefits under bandwidth-bound conditions.

In contrast, FlatAttention-based implementations substantially reduce HBM access time relative to \textit{FA-3}, primarily due to the lower I/O complexity of the FlatAttention dataflow, as described in \Cref{subsec:flat}. However, \textit{FlatSC}, which relies on the naive \textit{SW.Seq} collective implementation, incurs significant on-chip communication overhead, ultimately resulting in worse performance than FlashAttention-based approaches.

Replacing \textit{SW.Seq} with the tree-based software collective strategy in \textit{FlatTC} mitigates part of this overhead, yet on-chip inter-tile communication still accounts for more than 65\% of the total runtime across all evaluated prefill-phase \gls{mha} layers. In contrast, enabling efficient collective primitives in the \gls{noc} fabric (\textit{FlatHC}) substantially accelerates inter-tile communication, allowing FlatAttention to outperform FlashAttention across most \gls{mha} configurations.

Furthermore, the \textit{FlatAsync} implementation demonstrates additional performance gains by overlapping Softmax computation, data movement, and matrix multiplication. Overall, our proposed optimizations achieve up to a \ResAOptFlatPerfSU$\times$ speedup and a \ResAOptFlatHBMTraficSave$\times$ reduction in HBM traffic over \textit{FA-3} (D128, S4096).

\subsection{Tile Group Scale Trade-offs for FlatAttention}
\label{subsec:exp2}

Although \textit{FlatAsync} delivers the best performance across all prefill-phase \gls{mha} layers in \Cref{plot1}, it does not fully utilize the accelerator for shorter sequence lengths (e.g., $S=2048$ and $S=1024$), where the RedMulE execution cannot be completely overlapped with other pipeline stages. To identify the optimal FlatAttention configuration, we study the impact of different (square) group sizes, $G_x, G_y \in \{4, 8, 16, 32\}$, on the \arch accelerator configuration summarized in \Cref{tab:system_specs1}. We evaluate multiple \gls{mha} layers with sequence lengths $S \in \{512, 1024, 2048, 4096\}$, while fixing the head dimension to $D=128$, the number of heads to $H=32$, and the batch size to $B=4$.

\begin{figure}[t!]
  \centering
  \includegraphics[width=\columnwidth]{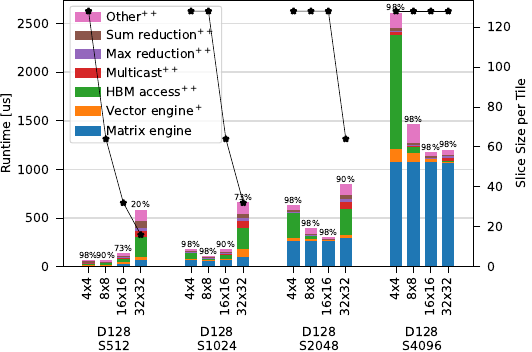}
  \caption{Runtime breakdown for different (square) flattening scales and layer sizes. Percentage labels above the bars indicate the average utilization of the matrix engine when active. $^+$Runtime not overlapped with matrix engine. $^{++}$Runtime not overlapped with either vector or matrix engine.
  }
  \label{plot2}
\end{figure}

\Cref{plot2} reports the runtime breakdown and the per-tile \gls{mha} workload slice size, i.e., $\frac{B_r}{G_y} = \frac{B_c}{G_x}$, across different group scales. For long sequences (e.g., $S=4096$), the per-tile slice size is bounded by L1 memory capacity and thus remains constant. In this regime, increasing the group size reduces the overall HBM I/O complexity, as discussed in \Cref{subsec:flat}, leading to shorter HBM access times and improved overlap with matrix multiplication on RedMulE. As a result, the $16 \times 16$ and $32 \times 32$ group configurations achieve \ResBUtia\% and \ResBUtib\% utilization, respectively, for $S=4096$.

In contrast, for shorter sequence lengths (e.g., $S=512$), increasing the group scale reduces the per-tile slice size due to the fixed total sequence length, introducing two sources of performance degradation:
\begin{itemize}
    \item \textbf{Reduced RedMulE utilization}: Smaller per-tile slices lead to lower compute efficiency. For example, under a $32 \times 32$ grouping with $S=512$, the workload slice assigned to every tile amounts to $\frac{B_r}{G_y} = \frac{B_c}{G_x} = 16$, and the RedMulE engines achieve only 20\% utilization during their active periods.

    \item \textbf{Amplified synchronization overhead}: Smaller slices shorten RedMulE execution time, causing fixed costs—such as synchronization and data movement, including \gls{hbm} access latency (approximately 200 cycles)—to account for a larger fraction of total runtime. Consequently, RedMulE execution can no longer fully hide these overheads, resulting in degraded performance.
\end{itemize}

We refer to this phenomenon as \textit{over-flattening}. For moderate sequence lengths, these effects coexist: larger group sizes reduce I/O complexity but simultaneously increase the risk of over-flattening, highlighting a trade-off between communication efficiency and compute utilization.

\begin{figure}[t!]
  \centering
  \includegraphics[width=\linewidth]{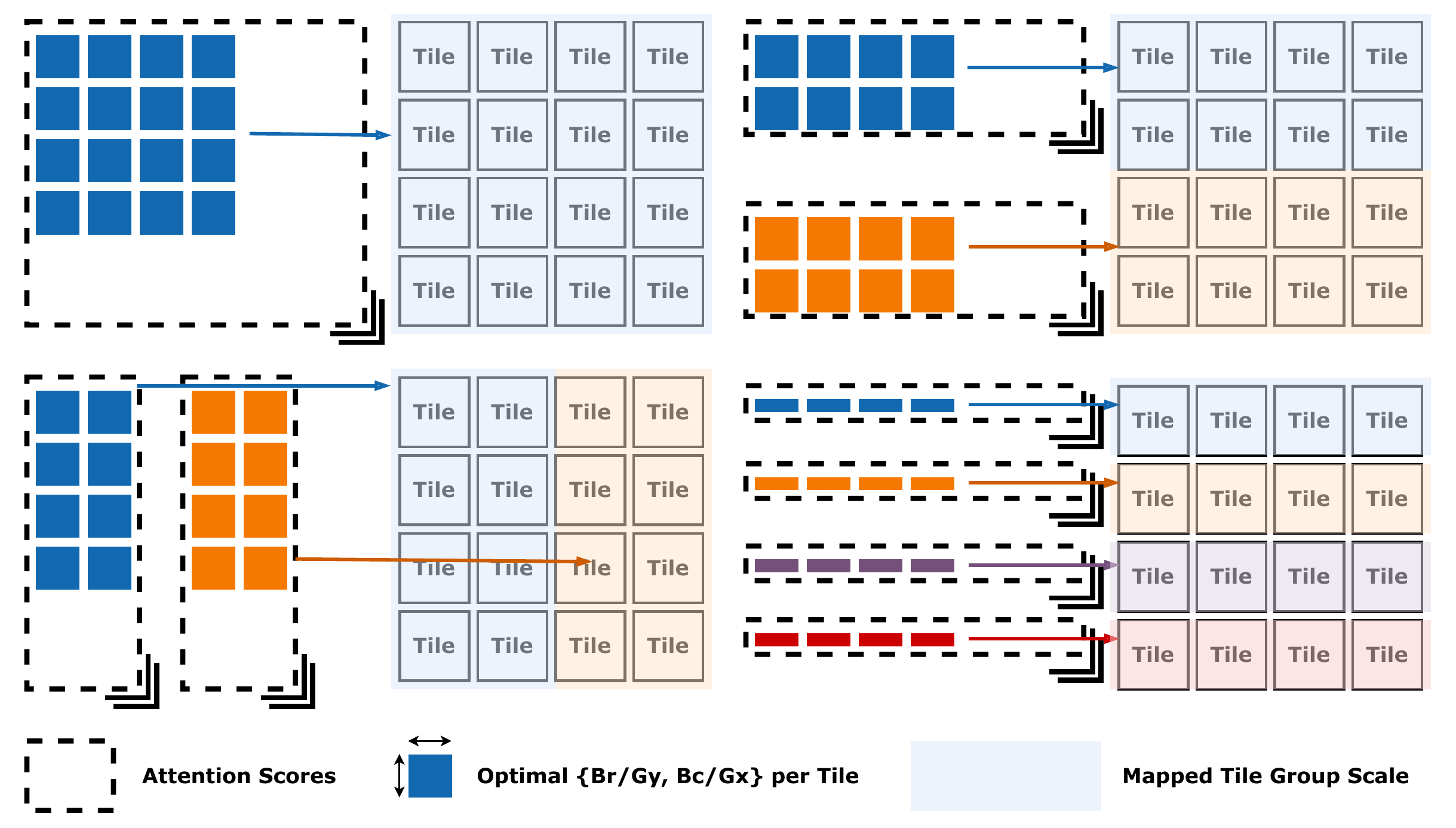}
  \caption{A general tiling and group-scaling strategy for FlatAttention.}
  \label{fig_tiling_strategy}
\end{figure}

\begin{figure}[t]
    \centering
    \begin{subfigure}{0.49\columnwidth}
        \centering
        \includegraphics[width=\textwidth, trim={0.25cm 0.25cm 0.25cm 0.25cm}, clip]{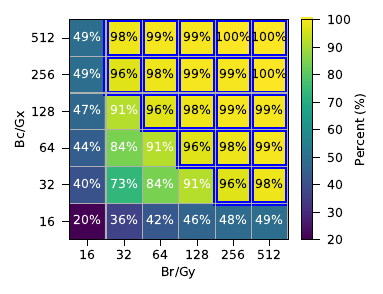}
        \put(-120,6){\small (a)}
        \captionsetup{labelformat=empty}
        \caption{}
        \vspace{-1em}
        \label{plot_RedMule_Matrix}
    \end{subfigure}
    \hfill
    \begin{subfigure}{0.49\columnwidth}
        \centering
        \includegraphics[width=\textwidth, trim={0.25cm 0.25cm 0.25cm 0.25cm}, clip]{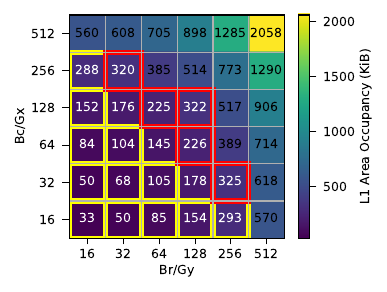}
        \put(-120,6){\small (b)}
        \captionsetup{labelformat=empty}
        \caption{}
        \vspace{-1em}
        \label{plot_L1_Needs}
    \end{subfigure}
    \vspace{-1.6em}
    \caption{(a) RedMulE utilization vs. tiling size (b) L1 area occupancy of \textit{FlatAsync} dataflow vs. tiling size.}
    \label{plot_tiling_explore}
    \vspace{-0.5em}
\end{figure}

\begin{figure*}[t!]
  \centering
  \includegraphics[width=\linewidth]{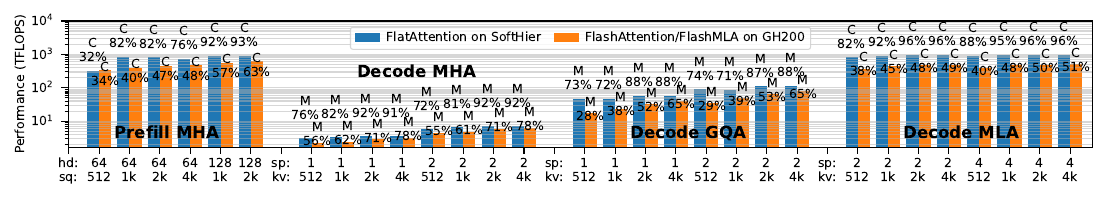}
  \vspace{-2em}
  \caption{Benchmark FlatAttention on \arch accelerator vs. optimized attention dataflow on Nvidia's GH200 GPU\cite{benchmarkresults}, varying head dimension (\textit{hd}) and sequence length (\textit{sq}) for prefill, speculative (\textit{sp}) and KV cache length (\textit{kv}) for decoding.}
  \label{plot_Flash_vs_Flat_All}
\end{figure*}

Motivated by the previous analysis, we propose a general tiling and group-scaling strategy for the FlatAttention dataflow that applies across attention varaints (\gls{mha}, \gls{gqa}, \gls{mla}). The key principle is to prioritize per-tile RedMulE utilization before aggressive flattening: we first determine an optimal per-tile tiling configuration, i.e., $\frac{B_r}{G_y}$ and $\frac{B_c}{G_x}$, that maximizes compute efficiency on each tile, and then increase the group scale as much as possible based on the shape of the attention score matrix and the mesh topology of the \arch accelerator. \Cref{fig_tiling_strategy} illustrates this strategy across different attention variants.

To identify the optimal per-tile tiling, we evaluate RedMulE utilization and L1 memory occupancy under varying $\frac{B_r}{G_y}$ and $\frac{B_c}{G_x}$ configurations, as shown in \Cref{plot_RedMule_Matrix} and \Cref{plot_L1_Needs}, respectively. Under the constraints of achieving more than 95\% RedMulE utilization while keeping L1 memory usage within the 384\,KiB budget specified in \Cref{tab:system_specs1}, we find that $\frac{B_r}{G_y} = \frac{B_c}{G_x} = 128$ is the optimal choice for the evaluated \arch configuration. This configuration delivers the highest RedMulE utilization, reaching up to \ResBRedMuleBestUti\%, without exceeding L1 capacity.

Following the proposed tiling and group-scaling strategy, we evaluate FlatAttention on the \arch accelerator across multiple attention variants, including prefill-phase \gls{mha}, decode \gls{mha}, decode \gls{gqa}, and decode \gls{mla}, under a wide range of input shapes.
We use the system configuration listed in \Cref{tab:system_specs1}, placing two HBM4 stacks on the south edge with up to 4 TB/s of bandwidth, which matches the peak FP16 performance and off-chip bandwidth of the NVIDIA GH200 GPU.
We compare FlatAttention on the said \arch accelerator against state-of-the-art attention implementations on GH200: FlashAttention for \gls{mha} and \gls{gqa}, and FlashMLA for \gls{mla}.

\Cref{plot_Flash_vs_Flat_All} summarizes the comparison results.
Bars labeled \texttt{C:x\%} denote compute-bound kernels (\texttt{x\%} utilization), while \texttt{M:y\%} denote memory-bound kernels (\texttt{y\%} \gls{hbm} bandwidth utilization).
Across all attention variants and input shapes, FlatAttention achieves an average utilization of \ResBFlatAvgUti\% in compute-bound regimes and an average \gls{hbm} bandwidth utilization of \ResBFlatAvgBWUti\% in memory-bound regimes.
Overall, FlatAttention on the \arch accelerator outperforms the optimized attention implementations on GH200 in most evaluated scenarios, delivering an average performance speedup of \ResBFlatGPUSU$\times$.

Beyond the FlatAttention performance analysis presented above for a given \arch configuration, FlatAttention can also be leveraged as a feedback mechanism to guide \arch accelerator design parameters. The detailed co-exploration of architectural parameters is presented in Appendix D.

\subsection{DeepSeek-v3 Decoder Inference Performance Analyze on Wafer-Scale Multi-Die System}
\label{subsec:exp4}

In this section, we shift our focus to end-to-end inference performance of DeepSeek-v3 on wafer-scale multi-die systems.
The system integrates 64 \arch accelerators interconnected via a wafer-scale interposer in an $8\times8$ mesh topology.
Each accelerator is connected through \gls{d2d} links with 1~TB/s bandwidth and 256~ns latency.
All \arch accelerators are configured with identical peak FP8 throughput (1976~TFLOPS without sparsity) and \gls{hbm} bandwidth (4~TB/s), as DeepSeek-v3 inference operates entirely in FP8 precision.

In the RedMulE matrix engine, FP8 peak throughput matches that of FP16.
To achieve this FP8 peak performance, we increase the accelerator operating frequency to 1.9~GHz from the previous configuration in \Cref{tab:system_specs1}.
Similar frequencies are achieved in the NVIDIA GH200 GPU.
Each accelerator is equipped with two HBM4 stacks, providing up to 4 TB/s of bandwidth and 128 GiB of HBM capacity, enabling full deployment of DeepSeek-v3-671B decoder inference with \gls{mtp} enabled on the 64-accelerator wafer-scale system.

\begin{figure*}[t]
    \centering
    \begin{subfigure}{0.5\columnwidth}
        \centering
        \includegraphics[width=\textwidth]{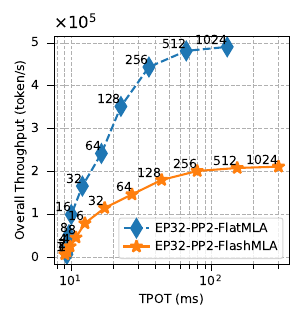}
        \put(-120,7.5){\small (a)}
        \captionsetup{labelformat=empty}
        \caption{}
        \vspace{-1.3em}
        \label{plot_DSv3_Flat_vs_Flash}
    \end{subfigure}
    \begin{subfigure}{0.45\columnwidth}
        \centering
        \includegraphics[width=\textwidth]{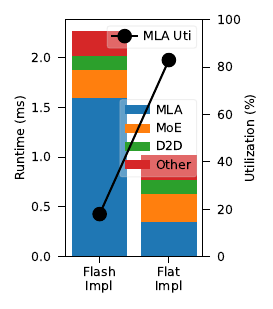}
        \put(-90,7){\small (b)}
        \captionsetup{labelformat=empty}
        \caption{}
        \vspace{-1.3em}
        \label{plot_DSv3_Flat_vs_Flash_breakdown}
    \end{subfigure}
    \begin{subfigure}{0.6\columnwidth}
        \centering
        \includegraphics[width=\textwidth]{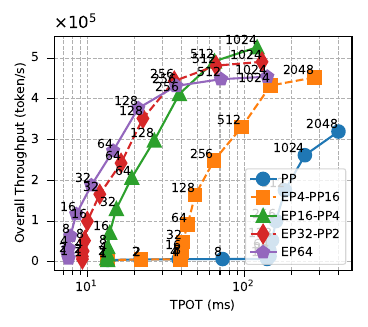}
        \put(-130,7){\small (c)}
        \captionsetup{labelformat=empty}
        \caption{}
        \vspace{-1.3em}
        \label{plot_DSv3_Decode_EP}
    \end{subfigure}
    \begin{subfigure}{0.35\columnwidth}
        \centering
        \includegraphics[width=\textwidth]{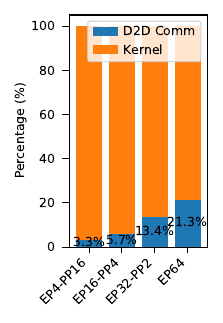}
        \put(-10,7){\small (d)}
        \captionsetup{labelformat=empty}
        \caption{}
        \vspace{-1.3em}
        \label{plot_C2COverhead}
    \end{subfigure}
    \vspace{-0.6em}
    \caption{(a) DeepSeek-v3-671B decoding performance on 64-accelerator system comparing using FlatAttention and FlashMLA dataflow, as well as (b) runtime breakdown for decoding with 256 batch size per chip. Effect of expert parallelism: (c) Performance on the wafer-scale system; (d) \gls{d2d} communication overhead at batch size of 256.}
    \vspace{-0.5em}
\end{figure*}

We first evaluate the impact of the FlatAttention dataflow on end-to-end decoding performance.
Specifically, we adopt the EP32--PP2 parallelization scheme, as described in \Cref{subsec:dsv3sys}.
By varying the number of batched user token requests per accelerator ($b$), we plot the system-level throughput versus \gls{tpot}, as shown in \Cref{plot_DSv3_Flat_vs_Flash}, comparing FlatAttention against FlashMLA.

At low batch sizes, \gls{tpot} is minimized while overall system throughput is low.
In this regime, FlatAttention and FlashMLA exhibit comparable performance.
As the batch size per accelerator increases, FlatAttention achieves up to $2.1\times$ higher system throughput than FlashMLA, while simultaneously delivering lower \gls{tpot}.

Figures~\ref{plot_DSv3_Flat_vs_Flash_breakdown} presents runtime breakdowns of one decoding layer at a batch size of 256 tokens per accelerator (within the \gls{tpot} constraint of 50 ms), using FlatAttention and FlashMLA, respectively.
In both cases, attention computation dominates decoding runtime, accounting for 42\% of total runtime with FlatAttention and 71\% with FlashMLA.
FlashMLA exhibits low accelerator utilization, whereas FlatAttention increases utilization to 83\%, resulting in a $4.5\times$ speedup for the attention component and translating to an end-to-end decoding speedup of $2.1\times$.

We further study the impact of expert parallelism on decoding performance, as shown in \Cref{plot_DSv3_Decode_EP}.
Under fully pipelined parallelism, increasing batch size in the low-batch regime does not improve overall throughput and instead degrades \gls{tpot}, primarily because not all experts are activated.
Once the batch size exceeds 16, at which point all experts become active, overall throughput increases, albeit with higher \gls{tpot}.

As expert parallelism scales from none to full, both system throughput and \gls{tpot} improve for low to medium batch sizes.
However, at high batch sizes, chip-to-chip communication overhead becomes a dominant contributor to decoding latency.
Increasing the degree of expert parallelism further amplifies this overhead, as illustrated in \Cref{plot_C2COverhead}, particularly in mesh-based accelerator interconnection topologies where multi-hop communication is unavoidable.
Future work should explore more optimized yet physically realizable interconnect topologies for wafer-scale systems.

Finally, we compare the DeepSeek-v3-671B decoding performance of \syspl with \gls{soa} serving systems built from commercial GPUs and NPUs.
\Cref{tab: dsv3_soa} summarizes the evaluated systems, including the number of chips, inter-chip interconnect topology, and decoding performance measured in per-chip throughput.
The compared wafer-scale systems show similar per-chip capabilities, particularly to the NVIDIA H800 used in \textit{DS-Prof}\cite{deepseek_profile}.
We report results under a decoding operating point that satisfies a 50\,ms \gls{tpot} constraint.

Compared to the best-performing \gls{soa} system (\textit{DS-Prof}), our \sys with 1~TB/s \gls{d2d} link bandwidth achieves a \ResDDSSoAPerfSUNvGPU$\times$ improvement in per-chip throughput and a \ResDDSSoATpotSUNvGPU$\times$ reduction in \gls{tpot}.
Even when the \gls{d2d} link bandwidth is reduced to a level comparable to the NVLink bandwidth used in the \textit{DS-Prof} system, our \sys continues to outperform \textit{DS-Prof}.
Despite the larger diameter of the 2D mesh wafer-scale interconnect and the lack of kernel–communication overlap, our \sys achieves a $1.6\times$ decoding throughput speedup.
This improvement is primarily driven by the increased utilization of attention kernels enabled by the FlatAttention dataflow proposed in this work.

In terms of overall system performance, DeepSeek-v3-671B decoding with FlatAttention on \syspl continues to outperform \textit{DS-Prof}, achieving up to a \ResDDSSoAPerfSUDSProf$\times$ improvement despite operating with \ResDDSSoAPeakRedDSProf$\times$ lower peak system performance.

\begin{table*}[!t]
    \centering
    \caption{DeepSeek-v3-671B Decoding Performnace Comaprision with \gls{soa} GPU/NPU Solutions}
    \label{tab: dsv3_soa}
    \renewcommand{\arraystretch}{1.2} 
    \setlength{\tabcolsep}{4pt} 
    \begin{tabular}{ccccccccc}
        \toprule
        \multirow{2}{*}{\textbf{System}} & \multirow{2}{*}{\textbf{No.Chips}} & \multirow{2}{*}{\textbf{Interconnect Meth}} & \multicolumn{5}{c}{\textbf{Per Chip}} & \multirow{2}{*}{\textbf{TPOT (ms)}}\\
        \cmidrule(lr){4-8}
         & & & \textbf{\gls{hbm}} & \textbf{TFLOPS} & \textbf{Batch} & \textbf{KV.Len} & \textbf{Token/s}\\ 
        \midrule
        CM384\cite{cloudmatrix384} & 384 Acsend 910C & Multi-Plane: UBLink 382GB/s, RDMA 400Gbps & 3.2 TB/s & 1504@INT8 & 128 & 4096 & 1943 & 49.4 \\
        \midrule
        DS-Prof\cite{deepseek_profile} & 96 Nvidia H800 & Multi-Plane: NV-Link 160GB/s, RDMA 400Gbps & 3.6 TB/s & 1979@FP8 & 128 & 4096 & 2325 & 50.2 \\
        \midrule
        Ours1 & 64 Tile Accel. & Wafer-Scale: 8$\times$8 Mesh, D2DLink 1TB/s & 4.0 TB/s & 1979@FP8 & 256 & 4096 & 6940 & 35.8 \\
        \midrule
        Ours2 & 64 Tile Accel. & Wafer-Scale: 8$\times$8 Mesh, D2DLink 160GB/s & 4.0 TB/s & 1979@FP8 & 128 & 4096 & 3773 & 33.1 \\
        \bottomrule
    \end{tabular}
    \vspace{-4mm}
\end{table*}

\section{Related Work}

\begin{table}[!t]
    \centering
    \caption{Related Work Comparison}
    \begin{threeparttable}
        \setlength{\tabcolsep}{2pt}
        \begin{tabular}{cccccccc}
            \toprule
            {\makecell{Works}} & {\makecell{Layer\\ Fusion}} & {\makecell{Atten-\\tion}} & {\makecell{Multiple\\Tiles}} & {\makecell{Archi-\\tecture}} & {\makecell{Coll-\\ectives}} & {\makecell{HW Mcast \\ / Redu.}} \\
            \midrule
            \cite{kao2023flat, nayak2024fusemax, shakerdargah2025masattention, lin2025systolicattention} & \cmark & \cmark & \xmark & - & \xmark & \xmark \\
            FlashAttention-2\cite{dao2023} & \cmark & \cmark & \cmark & gpu & \xmark & \xmark \\
            FlashFuser \cite{huang2025flashfuser} & \cmark & \xmark & \cmark & gpu & \cmark & \xmark \\
            Zen-Attention \cite{deshmukh2025zenattention} & \cmark & \cmark & \cmark & mesh & \cmark & \xmark \\
            COMET \cite{negi2025comet} & \cmark & \cmark & \cmark & noc & \cmark & \xmark \\
            ClusterFusion \cite{luo2025clusterfusion} & \cmark & \cmark & \cmark & gpu & \cmark & \xmark \\
            WaferLLM \cite{he2025waferllm} & \xmark* & \cmark & \cmark & mesh & \cmark & \cmark/\xmark \\
            \textbf{FlatAttention [Ours]} & \cmark & \cmark & \cmark & mesh & \cmark & \cmark \\
            \bottomrule
        \end{tabular}
        \begin{tablenotes}
            \footnotesize
            \item[*]Wafer-scale assumption: layer fusion is unnecessary for (small) models that fit entirely in on-chip memory.
        \end{tablenotes}
    \end{threeparttable}
    
    \label{tab:related_work}
    \vspace{-4mm}
\end{table}

A large body of works have developed optimized Attention dataflows, aiming to eliminate the costly off-chip transfers that arise when Softmax normalization is implemented as a separate stage from the surrounding \glspl{gemm} \cite{kao2023flat,nayak2024fusemax,shakerdargah2025masattention,lin2025systolicattention}.
The general solution is to \emph{fuse} all of the elementary operations in the Attention kernel to keep intermediate tensors on-chip.

While previous works focus on \emph{single}-tile (or \emph{single}-SM) accelerators, FlashAttention-2 \cite{dao2023} introduced a state-of-the-art fused Attention dataflow for \glspl{gpu}, achieving high efficiency through optimized tiling and work partitioning \textit{across} \glspl{sm}.
However, as inter-\gls{sm} communication on classic pre-Hopper \glspl{gpu} occurs through global memory, hindering layer fusion and making on-chip \gls{sm}-to-\gls{sm} communication prohibitively expensive, FlashAttention-2 adopts an embarassingly-parallel dataflow that explicitly avoids such communication.

Only in their most recent \gls{gpu} architecture, Hopper, Nvidia has introduced efficient \gls{sm}-to-\gls{sm} communication within \glspl{gpc} \cite{luhnen2024}, enabling efficient on-chip collective communication operations between \glspl{sm} in a \gls{gpc}.
Huang et al. \cite{huang2025flashfuser} have recently demonstrated that leveraging on-chip collective communication operations on Hopper \glspl{gpu} can expand the opportunities for layer fusion beyond a single \gls{sm}.
While this work is a first attempt at employing on-chip collectives in \gls{llm} workloads on \glspl{gpu}, it exclusively targets \gls{gemm} chains and \gls{ffn} layers.

In contrast, a few very recent works have started to explore multi-tile Attention dataflows leveraging on-chip collective communication operations, on \glspl{gpu} and other tile-based architectures \cite{deshmukh2025zenattention,negi2025comet,luo2025clusterfusion}.
Zen-Attention \cite{deshmukh2025zenattention}, develops a tiling framework for fully-fused Attention layers on tile-based AMD NPUs, leveraging on-chip spatial reductions.
Unfortunately, due to the scarcity of details on their fusion scheme and dataflow implementation, we are unable to directly compare their implementation to our work. 
Similarly, COMET \cite{negi2025comet} proposes a representation and modeling framework to evaluate the cost of on-chip collective communication in \gls{noc}-based accelerators for the exploration of compound operator dataflows, with a particular focus on Attention.
While the COMET representation appears to be generic enough to cover the FlatAttention dataflow at a high-level, the work proposes and analyzes collectives as a schedule-level cost rather than as an architecture-dataflow co-design knob.
On the other hand, ClusterFusion \cite{luo2025clusterfusion} proposes an optimized fully-fused Attention dataflow leveraging collective communications on Hopper \glspl{gpu}.
While their dataflow shares some similarities with ours, published earlier \cite{zhang2025}, it is not as flexible: ``flattening'' only along one axis.
Furthermore, architectural differences impose different tradeoffs and constraints.
For example, though they briefly investigate the tradeoffs involved in selecting an optimal ``flattening scale'' (in their case determined by the thread block cluster size), as their work mentions, the scope of this exploration is bounded by the size of a single \gls{gpc}.

Moreover, no prior work evaluates the use of hardware-accelerated collectives, with the exception of WaferLLM \cite{he2025waferllm}, which leverages hardware multicast support in Cerebras' WSE-2 accelerator's \gls{noc}.
However, as WaferLLM targets wafer-scale accelerators, it assumes the target models to fit entirely in on-chip memory, removing the need to develop a fused-layer dataflow.
As a consequence, their dataflow is not suited if they do not fit entirely within on-chip memory, and may fall back to a suboptimal Attention implementation with larger models, even on wafer-scale accelerators.
\Cref{tab:related_work} summarizes comparative analysis.
Note, none of the previous works comprehensively explore the challenge of co-designing an optimized tile-based accelerator for \glspl{llm}.

\section{Conclusion}

In this work, we propose \textit{FlatAttention}, a dataflow co-designed with fabric-level collective communication primitives supported by the \arch accelerator to efficiently accelerate modern attention variants in \glspl{llm}.
including \gls{mha}, \gls{gqa}, and \gls{mla}.
FlatAttention achieves an average \ResBFlatAvgUti\% utilization for compute-bound attention kernels and \ResBFlatAvgBWUti\% \gls{hbm} bandwidth utilization for memory-bound kernels, resulting in an average \ResBFlatGPUSU$\times$ speedup over optimized attention implementations on GH200.
We further evaluate end-to-end DeepSeek-v3 FP8 decoding on a \sys composed of 64 tile-based accelerators.
FlatAttention improves both system throughput and per-user latency,
achieving up to a \ResDDSSoAPerfSUDSProf$\times$ improvement in system throughput and a \ResDDSSoATpotSUNvGPU$\times$ reduction in \gls{tpot}, despite operating with \ResDDSSoAPeakRedDSProf$\times$ lower peak system performance compared to the \gls{soa} solution deployed on 96 NVIDIA H800 GPUs. 

\section*{Acknowledgments}
\noindent This work was supported by the ETH Future Computing Laboratory (EFCL), financed by a donation from Huawei Technologies.

\bibliographystyle{IEEEtran}
\bibliography{paper.bib}


\begin{IEEEbiography}[{\includegraphics[width=1in,height=1.25in,keepaspectratio,clip]{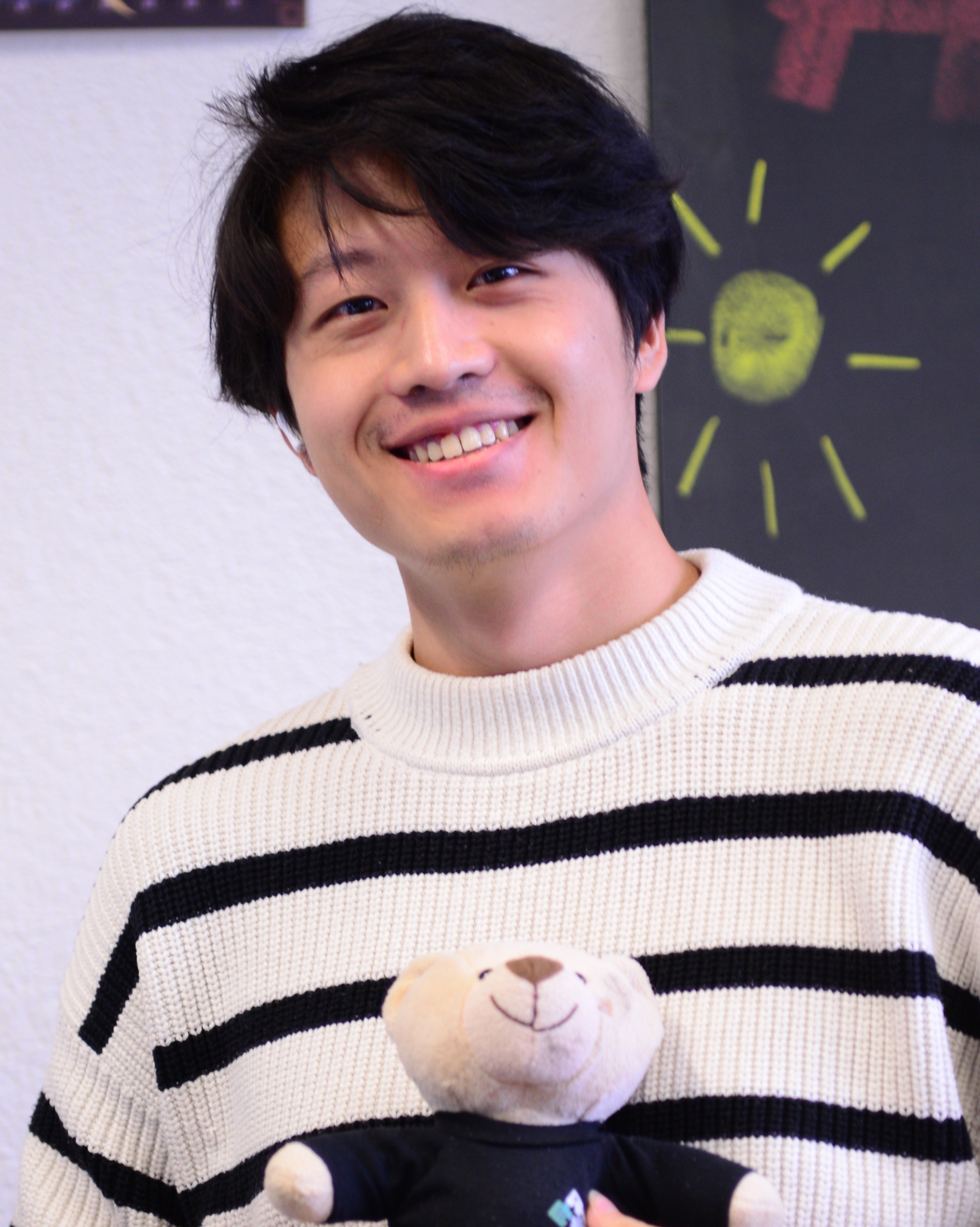}}]{Chi Zhang}
received his B.Sc. degree from Huazhong University of Science and Technology China in 2019 and his M.Sc. degree from KTH Royal Institute of Technology Sweden in 2022. He is currently pursuing a Ph.D. degree in the Digital Circuits and Systems group of Prof. Benini. His research interests include high-performance computing, memory systems, and near-memory computing.
\end{IEEEbiography}
\vspace{-15em}

\begin{IEEEbiography}[{\includegraphics[width=1in,height=1.25in,keepaspectratio,clip]{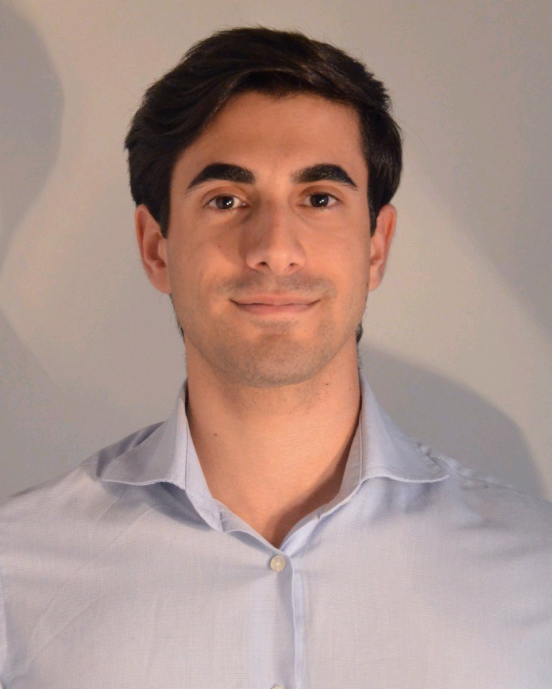}}]{Luca Colagrande}
received his BSc degree from Politecnico di Milano in 2018 and his MSc degree from ETH Zurich in 2020.
He is currently pursuing a PhD in the Digital Circuits and Systems group of Prof. Benini.
His research focuses on the co-design of energy-efficient general-purpose manycore accelerators for machine learning and high-performance computing applications.
\end{IEEEbiography}
\vspace{-15em}

\begin{IEEEbiography}[{\includegraphics[width=1in,height=1.25in,keepaspectratio,clip]{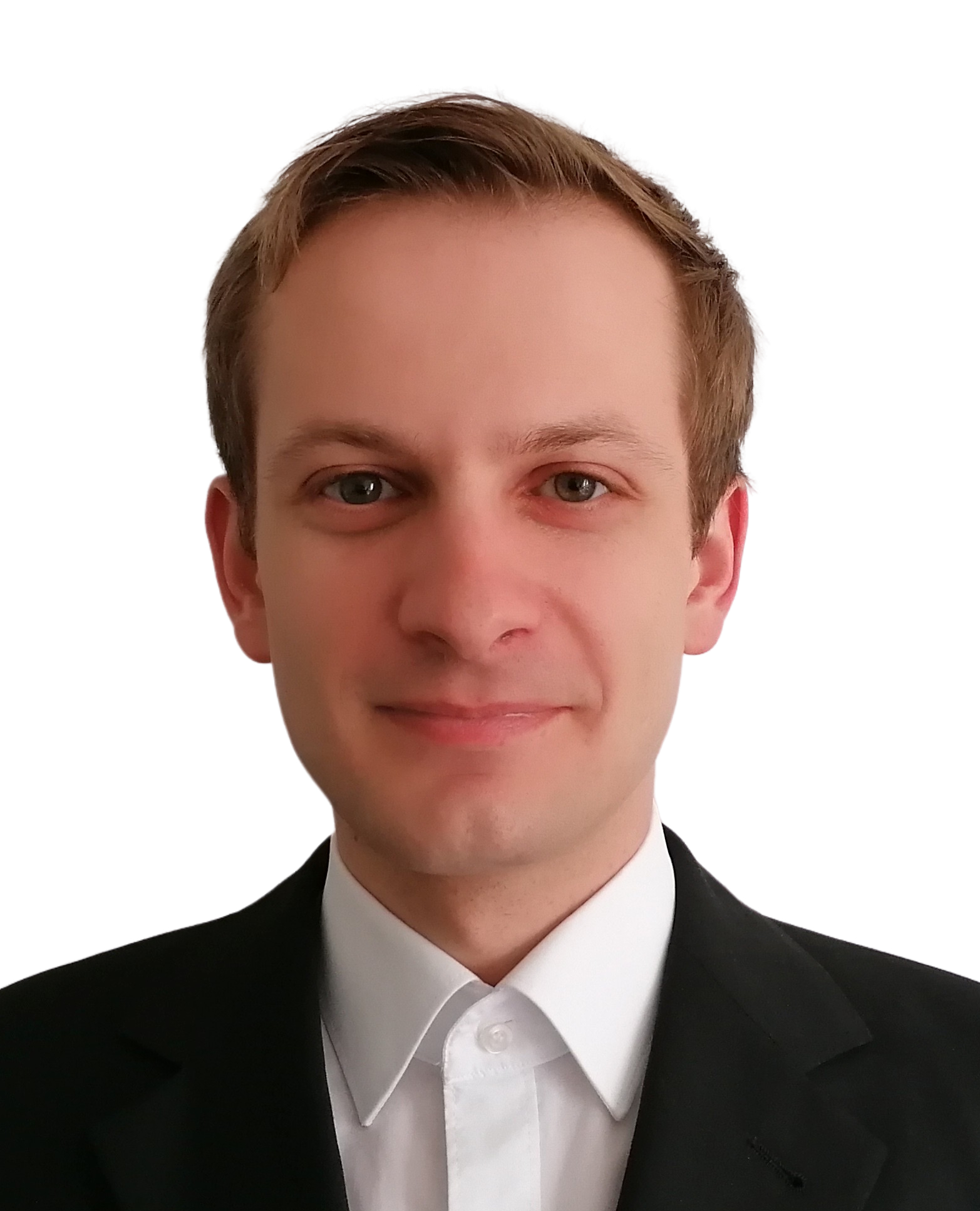}}]{Renzo Andri}
received the B.Sc., M.Sc. and Ph.D. degree in Electrical Engineering and Information Technology at ETH Zurich in 2013, 2015, and 2020, respectively. He is a principal researcher at the Computing Systems Laboratory, Huawei Technologies, Switzerland. His research interests are in energy-efficient machine learning hardware architecture. In 2019, he won the IEEE TCAD Donald O. Pederson Award.
\end{IEEEbiography}
\vspace{-15em}

\begin{IEEEbiography}[{\includegraphics[width=1in,height=1.25in,keepaspectratio,clip]{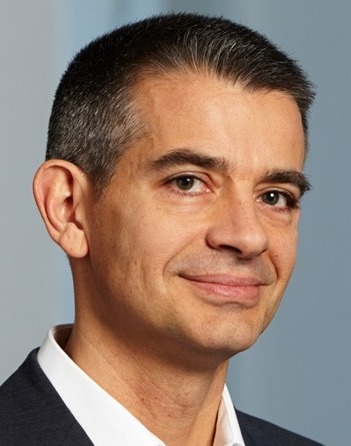}}]{Luca Benini}
holds the chair of Digital Circuits and Systems at ETH Zurich and is Full Professor at the Università di Bologna. He has served as Chief Architect for the Platform2012 in STMicroelectronics, Grenoble. Dr. Benini’s research interests are in energy-efficient parallel computing systems, multi-core SoC design, smart sensing micro-systems and machine learning hardware. He is a Fellow of the ACM and a member of the Academia Europaea.
\end{IEEEbiography}

\newpage

\vfill

\end{document}